\documentclass[twocolumn,showpacs,reprint,amsmath,amssymb,eqsecnum,pre]{revtex4-1}

\usepackage{graphicx}
\usepackage{dcolumn}
\usepackage{bm}

 \usepackage{color}

\usepackage{amssymb}
\usepackage{amsmath}
\usepackage{graphicx}
\usepackage{amsbsy}

\newcommand{\bq}{\begin{eqnarray}}
\newcommand{\eq}{\end{eqnarray}}
\newcommand{\bqn}{\begin{eqnarray*}}
\newcommand{\eqn}{\end{eqnarray*}}

\newcommand\beq{\begin{equation}}
\newcommand\eeq{\end{equation}}
\newcommand\beqa{\begin{eqnarray}}
\newcommand\eeqa{\end{eqnarray}}
\newcommand{\nn}{\nonumber\\}
\newcommand{\PY}{\text{PY}}
\newcommand{\zero}{{(0)}}
\newcommand{\one}{{(1)}}
\newcommand{\deltak}{\sigma^{\text{add}}}
\newcommand{\add}{\text{add}}
\newcommand{\aA}{\text{RFA}}
\newcommand{\aB}{\text{RFA}_+}
\newcommand{\s}{\tau}

\begin{document}
\title{Nonadditive hard-sphere fluid mixtures: A simple analytical theory}

\author{Riccardo Fantoni}
\email{rfantoni@ts.infn.it}
\homepage{http://www-dft.ts.infn.it/~rfantoni/}
\affiliation{National Institute for Theoretical Physics (NITheP) and
  Institute of Theoretical Physics, Stellenbosch 7600, South Africa}

\author{Andr\'es Santos}
\email{andres@unex.es}
\homepage{http://www.unex.es/eweb/fisteor/andres}
\affiliation{Departamento de F\'isica, Universidad de Extremadura,
  E-06071 Badajoz, Spain}

\date{\today}

\begin{abstract}
We construct a non-perturbative fully analytical approximation for
the thermodynamics and the structure of non-additive hard-sphere fluid mixtures. The method  essentially lies in  a heuristic extension of
the Percus--Yevick solution for additive hard spheres. Extensive
comparison with Monte Carlo simulation data shows a generally good agreement, especially in the case of like-like radial distribution functions.
\end{abstract}


\maketitle
\section{Introduction}
\label{sec:introduction}

The  van der Waals ideas \cite{HM06} show that the most
important feature of
the pair potential between atoms or molecules is the harsh repulsion
that appears at short range and has its origin in the overlap of the
outer electron shells. These ideas form the basis of the very successful
perturbation theories of the liquid state. This, along with fruitful applications to soft matter \cite{L01}, explains the continued interest in hard-sphere reference systems \cite{M08}.

The simplest model for a
fluid \emph{mixture} is a system of
additive hard spheres (AHSs) for which the
like-unlike collision diameter ($\sigma_{ij}$)  between a particle of
species $i$ and one of species $j$ is equal to the arithmetic mean
$\deltak_{ij}\equiv\frac{1}{2}(\sigma_{ii}+\sigma_{jj})$. A more general model consists of
\emph{nonadditive} hard spheres (NAHSs), where the like-unlike
collision diameter differs from $\deltak_{ij}$ by a quantity
$\Delta_{ij}=(\sigma_{ij}-\deltak_{ij})/\deltak_{ij}$ called the nonadditivity parameter. As mentioned in the paper by Ballone et
al.\ \cite{BPGG86}, where
the relevant references may be found, experimental work on alloys,
aqueous electrolyte solutions, and molten salts suggests
that homocoordination and heterocoordination \cite{GPE89,GPF90}  may be interpreted
in terms of excluded volume effects due to nonadditivity (positive and negative, respectively)
of the repulsive part of the intermolecular potential. NAHS systems
are also useful models to describe real physical systems as rare gas
mixtures \cite{S89} and colloids \cite{GHR83,LPPSW92,DBE99,MF94}.
For a short review of the literature on NAHSs up to 2005 the reader is referred to Ref.\ \cite{SHY05}.

The well-known Percus--Yevick (PY) integral-equation theory \cite{HM06} is exactly solvable for a mixture of three-dimensional (3D) AHS mixtures \cite{L64,YSH98}. The solution has been recently extended to any odd dimensionality \cite{RS11}. On the other hand, any amount of nonadditivity ($\Delta_{ij}\neq 0$) suffices to destroy the analytical character of the solution and so one needs to resort to numerical methods to solve the PY or other integral equations \cite{BPGG86}.

The aim of the present paper is to propose a non-perturbative and fully analytical approach for 3D NAHS fluid mixtures, which can be seen as a na\"ive heuristic \emph{extension}  of the PY solution for AHS mixtures. In doing this, we are guided by the exact
solution of the one-dimensional (1D) NAHS model \cite{SZK53,LZ71,HC04,S07}
and   some physical
constraints are imposed: the radial distribution function (RDF) $g_{ij}(r)$ must be zero within
the diameter $\sigma_{ij}$, the isothermal compressibility must be
finite, and the zero density limit of the RDF
must be satisfied. We find that this strategy gives very good results both for the thermodynamics and the
structure, provided that some geometrical constraints on the diameters
and the nonadditivity parameter are satisfied.  This makes our approach
particularly appealing as a reference approximation for integral
equation theories and perturbation theories of fluids.

The paper is organized as follows: in Sec.\ \ref{sec:model} we describe
the NAHS model outlining the physical constraints that we want to
embody in our approach. The latter is constructed by a three-stage procedure (approximations $\aA$, $\aB$,  and $\aB^{(m)}$) in Sec.\ \ref{sec:approximation}. In
Sec.\ \ref{sec:EOS} we present the results for the equation of state
from our approximation, comparing them with available Monte Carlo (MC)
simulations. The results for the structural properties are presented in Sec.\ \ref{sec:structure}, where we compare  with our own MC simulations. Finally, Sec.\ \ref{sec:conclusions} is devoted to some concluding
remarks.

\section{The NAHS model}
\label{sec:model}

An $n$-component mixture of NAHSs in the
$d$-dimensional Euclidean space is a fluid of $N_i$ particles of
species $i$ (with $i=1,2,\ldots,n$), such that there are a total number
of particles $N=\sum_{i=1}^n N_i$ in a
volume $V$, and the pair potential between a particle of species $i$
and a particle of species $j$ separated by a distance $r$ is given by
\beq
U_{ij}(r)=\left\{\begin{array}{ll}
\infty, & r<\sigma_{ij}, \\
0   ,    & r>\sigma_{ij},
\end{array}\right.
\eeq
where $\sigma_{ii}=\sigma_i$ and
$\sigma_{ij}=\frac{1}{2}(\sigma_i+\sigma_j)(1+\Delta_{ij})$, so that
$\Delta_{ii}=0$ and $\Delta_{ij}=\Delta_{ji}>-1$. When
$\Delta_{ij}=0$ for all pairs $(i,j)$ we recover the AHS system. In a
binary mixture ($n=2$), $\Delta_{12}=\Delta_{21}=\Delta$ is the only  nonadditivity parameter. If $\Delta=-1$ one recovers the case of
two independent one-component hard-sphere (HS) systems.
In the other extreme case $\sigma_1=\sigma_2=0$ with $\sigma_{12}$
finite (so that $\Delta\to\infty$) one obtains the well known Widom--Rowlinson (WR) model
\cite{WR70,R71}. Another interesting case is the Asakura--Oosawa model \cite{AO54,AO58} (where $\sigma_2=0$ and $\Delta>0$), often used to discuss polymer colloid mixtures and
where the notion of a depletion potential was introduced. The NAHS system undergoes a demixing phase
transition  for positive nonadditivity
\cite{RP94,JY03,G03,B05,LALA96}. A demixing transition might also be possible, even for negative nonadditivity \cite{SH05,SH10}, provided the asymmetry ratio $\sigma_1/\sigma_2$ is sufficiently far from unity. In
the present paper we will only consider the NAHS system in its single fluid
phase.

Let the number density of the mixture be $\rho=N/V$ and the mole fraction
of species $i$ be $x_i=\rho_i/\rho$, where $\rho_i=N_i/V$ is the number
density of species $i$. From these quantities one can define the (nominal)
packing fraction $\eta=v_d\rho M_d$, where
$v_d=(\pi/4)^{d/2}\Gamma(1+d/2)$ is the volume of a $d$-dimensional
sphere of unit diameter and
\beq
M_k\equiv\langle\sigma^k\rangle=\sum_{i=1}^n x_i\sigma_i^k
\eeq
denotes the $k$th moment of the diameter distribution.

The NAHS model, in the thermodynamic limit $N\to\infty$ with $\rho\equiv N/V$
constant, admits an analytical exact solution for the structure and the
thermodynamics in $d=1$
\cite{SZK53,LZ71,HC04,S07}. Moreover, the AHS model in odd dimensions
 is
analytically solvable in the PY approximation \cite{L64,YSH98,RS11}, the result reducing to the exact
solution of the problem for $d=1$ but not for $d\geq 3$.

\subsection{Basic physical constraints on the structure}
\label{sec:constraints}
The RDF $g_{ij}(r)$ must comply with three basic conditions:
\begin{itemize}
\item[(i)] $g_{ij}(r)$  must vanish for $r<\sigma_{ij}$. More specifically, for distances near $\sigma_{ij}$,
\beq
g_{ij}(r)=\Theta(r-\sigma_{ij})\left[g_{ij}(\sigma_{ij}^+)+
g_{ij}^\prime(\sigma_{ij}^+)(r-\sigma_{ij})+\cdots\right],
\label{cond_i}
\eeq
where $\Theta(x)$ is the Heaviside step function.
\item[(ii)] In the fluid phase the isothermal compressibility $\chi$
must be finite. This implies (see below) that the Fourier
transform $\widetilde{h}_{ij}(q)$ of the total correlation function  $h_{ij}(r)\equiv g_{ij}(r)-1$ has to remain finite at $q=0$ or, equivalently,
\beq
\int_0^\infty dr\,r^\alpha h_{ij}(r)=\mbox{finite for $0\leq\alpha\le
d-1$}.
\label{cond_ii}
\eeq
\item[(iii)] In the low density limit, the RDF is
\beq
\lim_{\rho\to 0}g_{ij}(r)=e^{-U_{ij}(r)/k_BT}=
\Theta(r-\sigma_{ij}),
\label{cond_iii}
\eeq
$k_B$ and $T$ being the  Boltzmann constant and the
absolute temperature, respectively.

\end{itemize}
As a complement to Eq.\ \eqref{cond_iii}, we give below the exact expression of $g_{ij}(r)$ to first order in density \cite{YSH08}:
\beqa
g_{ij}(r)&=&\Theta(r-\sigma_{ij})\Big\{1+\frac{\pi\rho}{12r}\sum_{k=1}^n x_k \Theta(\sigma_{ik}+\sigma_{kj}-r)\nn
&&\times(r-\sigma_{ik}-\sigma_{kj})^2\left[r^2+2(\sigma_{ik}+\sigma_{kj})r\right.\nn
&&\left.-3(\sigma_{ik}-\sigma_{kj})^2\right]
+
\mathcal{O}(\rho^2)\Big\}.
\label{grho2exact}
\eeqa

\subsection{The two routes to thermodynamics}

For an athermal fluid like NAHSs there are two main routes that lead
from the knowledge of the structure to the equation of state
(EOS) \cite{HM06}. These may give different results for an
approximate  RDF.

The virial route to the EOS of the NAHS mixture
requires the knowledge of the contact values $g_{ij}(\sigma_{ij}^+)$
of the RDF,
\beq \label{Zv}
Z^v(\eta)=1+\frac{2^{d-1}}{M_d}\eta\sum_{i,j=1}^nx_ix_j\sigma_{ij}^d
g_{ij}(\sigma_{ij}^+),
\eeq
where $Z=p/\rho k_BT$ is the compressibility factor of the mixture,
$p$ being the pressure.

 The isothermal compressibility $\chi$, in a mixture, is in general given by
\beqa \label{chi}
\chi^{-1}&=&\frac{1}{k_BT}\left(
\frac{\partial p}{\partial\rho}\right)_{T,\{x_j\}}=
\frac{1}{k_BT}\sum_{i=1}^nx_i\left(
\frac{\partial p}{\partial\rho_i}\right)_{T,\{x_j\}}\nn
&=&
1-\rho\sum_{i,j=1}^nx_ix_j\widetilde{c}_{ij}(0),
\eeqa
where $\widetilde{c}_{ij}(q)$ is the Fourier transform of the
direct correlation function $c_{ij}(r)$, which is defined by the
Ornstein--Zernike (OZ) equation
\beq
\widetilde{h}_{ij}(q)=\widetilde{c}_{ij}(q)+\sum_{k=1}^n\rho_k
\widetilde{h}_{ik}(q)\widetilde{c}_{kj}(q).
\eeq
Introducing the quantities
$\widehat{h}_{ij}(q)\equiv \sqrt{\rho_i\rho_j}\widetilde{h}_{ij}(q)$ and
$\widehat{c}_{ij}(q)\equiv \sqrt{\rho_i\rho_j}\widetilde{c}_{ij}(q)$, the OZ
relation becomes, in matrix notation,
\beq
\widehat{\mathsf{c}}(q)=\widehat{\mathsf{h}}(q)\cdot
\left[\mathsf{I}+\widehat{\mathsf{h}}(q)\right]^{-1},
\eeq
where $\mathsf{I}$ is the $n\times n$ identity matrix. Thus
Eq. (\ref{chi}) can be rewritten as
\beqa
\chi^{-1}&=&\sum_{i,j=1}^n\sqrt{x_ix_j}
\left[\delta_{ij}-\widehat{c}_{ij}(0)\right]\nn
&=&
\sum_{i,j=1}^n\sqrt{x_ix_j}
\left[\mathsf{I}+\widehat{\mathsf{h}}(0)\right]^{-1}_{ij}.
\label{chih}
\eeqa
In Eq.\ \eqref{chih}, and henceforth, we use the notation $A_{ij}^{-1}$ to denote the $ij$ element of the inverse $\mathsf{A}^{-1}$ of a given square matrix $\mathsf{A}$.

In the particular case of binary mixtures ($n=2$), Eq.\ \eqref{chih} yields
\beq \label{chi_binary}
\chi=\frac{[1+\rho x_1 \widehat{h}_{11}(0)][1+\rho x_2 \widehat{h}_{22}(0)]
-\rho^2 x_1 x_2 \widehat{h}_{12}^2(0)}
{1+\rho x_1x_2 [\widehat{h}_{11}(0)+ \widehat{h}_{22}(0)-2 \widehat{h}_{12}(0)]}.
\eeq
The compressibility route to the EOS can be obtained from
\beq \label{Zc}
Z^c(\eta)=\int_0^1 {d x}\,{\chi^{-1}(\eta x)}.
\eeq

\subsection{The one-dimensional system\label{sec2C}}
The exact solution for nonadditive hard rods ($d=1$) is known \cite{LPZ62,S07,BNS09}. First, let us introduce the Laplace transform
\beq
G_{ij}(s)\equiv \int_0^\infty dr\, e^{-sr} g_{ij}(r).
\eeq
In terms of this quantity the exact solution has the form
\beq
G_{ij}(s)=\frac{1}{\sqrt{x_i x_j}}\sum_{k=1}^n {P}_{ik}(s)Q_{kj}(s),
\label{s1}
\eeq
where
\beq
{P}_{ij}(s)\equiv\sqrt{x_ix_j}
K_{ij}\frac{e^{-\sigma_{ij}(s+\xi)}}{s+\xi}
\label{s2}
\eeq
is proportional to the Laplace transform of the nearest-neighbor probability distribution and
\beq
\mathsf{Q}(s)\equiv \left[\mathsf{I}-\rho{\mathsf{P}}(s)\right]^{-1}.
\label{s3}
\eeq
In Eqs.\ \eqref{s1} and \eqref{s2}, $\xi\equiv p/k_BT=\rho Z$, while $K_{ij}=K_{ji}$  are state-dependent parameters that are determined as functions of $\xi$ from the condition \eqref{cond_ii}, which implies $\lim_{s\to 0} s G_{ij}(s)=1$, as well as requiring the ratio $K_{ij}/K_{ik}$ to be independent of $i$ \cite{S07}. Those conditions also provide the exact EOS in  implicit form, i.e., $\rho$ as a function of $\xi$.

Of course, the above results also hold for additive hard rods. In that case,
the additive property $\sigma_{ij}=\deltak_{ij}\equiv\frac{1}{2}(\sigma_i+\sigma_j)$
allows us to rewrite the solution in other equivalent ways. To that end,
let us define
\beq
L_{ij}=K_{ij}e^{-\xi\deltak_{ij}},
\label{1}
\eeq
so that
\beq
{P}_{ij}(s)=\sqrt{x_i x_j}L_{ij}\frac{e^{-\deltak_{ij}s}}{s+\xi},
\label{2}
\eeq
\beq
Q^{-1}_{ij}(s)=e^{a_{ij}s}\sqrt{\frac{x_j}{x_i}}\frac{s}{s+\xi}C_{ij}(s),
\label{2b}
\eeq
where
\beq
a_{ij}\equiv \frac{1}{2}(\sigma_i-\sigma_j)
\label{aij}
\eeq
and
\beq
C_{ij}(s)\equiv\left(1+\frac{\xi}{s}\right)\delta_{ij}-\frac{\rho
x_i}{s}L_{ij}e^{-\sigma_{i}s}.
\label{3}
\eeq
Here we have made use of the property
\beq
\sigma_i=\deltak_{ij}+a_{ij}.
\label{4}
\eeq
It is easy to prove that
\beq
Q_{ij}(s)=e^{a_{ij}s}\sqrt{\frac{x_j}{x_i}}\frac{s+\xi}{s}C^{-1}_{ij}(s),
\label{5}
\eeq
thanks to the property $a_{ik}+a_{kj}=a_{ij}$. Consequently, in the additive case, Eq.\
\eqref{s1} becomes
\beq
G_{ij}^{\text{add}}(s)=\frac{e^{-\deltak_{ij}s}}{s}\sum_{k=1}^n L_{ik}C^{-1}_{kj}(s),
\label{6}
\eeq
where use has been made of the additivity property
\beq
\deltak_{ik}-a_{kj}=\deltak_{ij}.
\label{7}
\eeq

The additive solution turns out to be
\beq
L_{ij}=\frac{\xi}{\rho}=\frac{1}{1-\rho M_1}.
\label{8}
\eeq
The fact that $L_{ij}=\text{const}$ allows one to rewrite Eq.\
\eqref{6} in yet another equivalent form,
\beq
G_{ij}^{\text{add}}(s)=\frac{e^{-\deltak_{ij}s}}{s}\sum_{k=1}^n L_{ik}B^{-1}_{kj}(s),
\label{9}
\eeq
where
\beqa
B_{ij}(s)&\equiv&\delta_{ij}-\frac{\rho
x_i}{s}L_{ij}\varphi_0(\sigma_{i}s)\nn
&=&C_{ij}(s)+\frac{\xi}{s}\left(x_i-\delta_{ij}\right).
\label{10}
\eeqa
In the first equality,
\beq
\varphi_0(x)\equiv e^{-x}-1.
\eeq
While  $\lim_{s\to 0}s C_{ij}(s)=\xi\left(\delta_{ij}- x_i \right)\neq 0$,
but $\det [s\mathsf{C}(s)]=\mathcal{O}(s)$, in the case of the matrix
$\mathsf{B}(s)$ one has $\lim_{s\to 0}s B_{ij}(s)=0$. On the other
hand, in both cases, $\lim_{s\to\infty} C_{ij}(s)=\lim_{s\to\infty}
B_{ij}(s)=\delta_{ij}$, so that $\lim_{s\to\infty} s
e^{\deltak_{ij}}G_{ij}(s)=L_{ij}=\xi/\rho$.

It turns out that Eqs.\ \eqref{6}, \eqref{8}, and \eqref{9} are also obtained from the PY solution for additive hard rods. Thus,  the PY equation yields the exact solution in the additive case, but
not in the nonadditive one.

It is important to bear in mind that, if one inverts the steps,
it is possible to formally get Eq.\ \eqref{s1} from Eq.\ \eqref{6}.  In other words, starting from the form \eqref{6} of the PY solution for the 1D AHS system, allowing $L_{ij}$ and $\xi$ to be free, and carrying out some formal manipulations, one arrives at an equivalent form, Eq.\ \eqref{s1}, that, if heuristically extended to the NAHS case ($\sigma_{ij}\neq \deltak_{ij}$), coincides with the exact solution to the problem. However,  it is not possible to recover \eqref{s1} starting from the form  \eqref{9} since the property $L_{ij}=\text{const}$, only valid in the additive case, cannot be reversed.

\subsection{PY solution for three-dimensional AHSs\label{sec2D}}
In this subsection we recall the PY solution for AHSs in three dimensions ($d=3$)  \cite{L64,YSH98}.

First, one introduces the Laplace transform of $r g_{ij}(r)$,
\beq
G_{ij}(s)\equiv \int_0^\infty dr\, e^{-sr} r g_{ij}(r).
\label{Gij}
\eeq
{}From Eq.\ (\ref{cond_i}) it follows that
\beqa
s e^{\sigma_{ij}s}G_{ij}(s)&=&\sigma_{ij}g_{ij}(\sigma_{ij}^+)+
\left[g_{ij}(\sigma_{ij}^+)+\sigma_{ij}g^\prime_{ij}(\sigma_{ij}^+)
\right]s^{-1}\nn
&&+\mathcal{O}(s^{-2}).
\label{slarge}
\eeqa
Next, Eq.\ \eqref{cond_ii} implies, for small $s$,
\beq
\label{s2G}
s^2G_{ij}(s)=1+H^{(0)}_{ij}s^2+H^{(1)}_{ij}s^3+\cdots
\eeq
with $H^{(0)}_{ij}=\text{finite}$ and
$H^{(1)}_{ij}=-\widetilde{h}_{ij}(0)/4\pi=\text{finite}$, where in general
\beq
H^{(\alpha)}_{ij}\equiv\frac{1}{\alpha!}\int_0^\infty dr\,(-r)^\alpha rh_{ij}(r).
\eeq
Finally, Eq.\ \eqref{cond_iii} yields
\beq
\lim_{\rho\to 0}G_{ij}(s)=\frac{e^{-\sigma_{ij}s}}{s^2}\left(1+\sigma_{ij}s\right).
\label{Grho}
\eeq
Equations \eqref{Gij}--\eqref{Grho} hold both for NAHSs and AHSs.

The PY solution for AHSs can then be written as \cite{L64,YSH98}
\beq \label{Grfa}
G_{ij}^{\text{add}}(s) = \frac{e^{-\deltak_{ij}s}}{s^2}\sum_{k=1}^n L_{ik}(s)B^{-1}_{kj}(s),
\eeq
where $\mathsf{L}(s)$ and $\mathsf{B}(s)$ are  matrices  given by
\beq \label{Lserie}
{L}_{ij}(s)= {L}_{ij}^\zero+{L}_{ij}^\one s,
\eeq
\beq
{B}_{ij}(s)= \delta_{ij}+\frac{2\pi\rho
 x_i}{s^3}\left[L_{ij}^\zero\varphi_2(\sigma_i s)+L_{ij}^\one s\varphi_1(\sigma_i s)\right],
 \label{Bdef}
\eeq
where
\beq
\varphi_1(x)\equiv e^{-x}-1+x,\quad \varphi_2(x)\equiv e^{-x}-1+x-\frac{x^2}{2}.
\eeq
Similarly to the 1D case, $\lim_{s\to 0}s B_{ij}(s)=0$. In fact, Eqs.\
 \eqref{Grfa}--\eqref{Bdef} are the 3D analogs of Eqs.\ \eqref{9} and
 \eqref{10}.
{}For the general structure of the PY solution with $d=\text{odd}$, the reader is referred to Ref.\ \cite{RS11}.

Also as in the 1D case,  $\lim_{s\to \infty} B_{ij}(s)=\delta_{ij}$ and
 so, according to Eq.\ \eqref{slarge},
\beq
\label{contact-value}
g_{ij}^\add(\sigma_{ij}^+)=\frac{L_{ij}^\one}{\deltak_{ij}}.
\eeq
Further, in view of Eq. (\ref{s2G}), the coefficients of $s^0$ and $s$
 in the power series expansion of $s^2G_{ij}(s)$ must be 1 and 0,
 respectively. This yields $2n^2$ conditions that allow us to find \cite{YSH98}
\beq
L_{ij}^{(0)}=\theta_1+\theta_2\sigma_j~,\quad L_{ij}^{(1)}=\theta_1\deltak_{ij}+\frac{1}{2}\theta_2\sigma_i\sigma_j,
\label{LijPY}
\eeq
where $\theta_1\equiv 1/(1-\eta)$ and $\theta_2\equiv
 3(M_2/M_3)\eta/(1-\eta)^2$.
It is straightforward to check that Eq.\ \eqref{Grfa} complies with the limit \eqref{Grho}.

The expressions (\ref{Zv}) and  (\ref{Zc}) which follow from the solution of the PY
equation of AHS mixtures are
\beq
Z_{\PY}^v(\eta)=\frac{1}{1-\eta}+\frac{M_1M_2}{M_3}\frac{3\eta}{(1-\eta)^2}
+\frac{M_2^3}{M_3^2}\frac{3\eta^2}{(1-\eta)^2},
\eeq
\beq
Z_{\PY}^c(\eta)=\frac{1}{1-\eta}+\frac{M_1M_2}{M_3}\frac{3\eta}{(1-\eta)^2}
+\frac{M_2^3}{M_3^2}\frac{3\eta^2}{(1-\eta)^3}.
\eeq
Usually, the virial route underestimates the exact results, while the compressibility route overestimates them.

\section{Construction of the approximations}
\label{sec:approximation}

As stated in Sec.\ \ref{sec:introduction}, the main aim of this paper is to construct  analytical approximations
for the structure and thermodynamics of 3D NAHSs. On the one hand, the approximations will be inspired by the exact solution in the 1D case (see Sec.\ \ref{sec2C}). On the other hand, they will reduce to the AHS PY solution (see Sec.\ \ref{sec2D}). Moreover, as a guide in the construction of the approximations and also to determine the parameters, the basic physical requirements \eqref{cond_i}--\eqref{cond_iii} [or, equivalently, \eqref{slarge}, \eqref{s2G}, and \eqref{Grho}] will be enforced.

The driving idea is to rewrite Eq.\ \eqref{Grfa} in a form akin to that of Eq.\
\eqref{s1}, by inverting the procedure followed in Sec.\ \ref{sec2C}. This method faces several difficulties. One of them is that, as said before,
Eq.\ \eqref{Grfa} is the 3D analog of Eq.\ \eqref{9}, but not of Eq.\ \eqref{6}, and it is not possible to recover directly (i.e., without further assumptions) Eq.\ \eqref{s1} from Eq.\ \eqref{9}.
One could first try to rewrite Eq.\ \eqref{Grfa} in a form akin to that of Eq.\ \eqref{6}, i.e., a form where the matrix $\mathsf{B}$ given by
Eq.\ \eqref{Bdef} is replaced by a matrix $\mathsf{C}$ such that $\lim_{s\to 0}s C_{ij}(s)\neq 0$. But, given the intricate structure of Eq.\ \eqref{Bdef} and the fact that neither $L_{ij}^\zero$ nor $L_{ij}^\one$ are constant, this does not seem to be an easy task at all. Therefore, we will work from Eq.\ \eqref{Grfa} directly.

\subsection{The AHS PY solution revisited}
First, define
\beq
{P}_{ij}(s)\equiv\sqrt{x_i x_j}e^{-\deltak_{ij}s}L_{ij}(s),
\label{11}
\eeq
\beq
{Q}_{ij}(s)\equiv e^{a_{ij}s}\sqrt{\frac{x_j}{x_i}}B^{-1}_{ij}(s),
\label{12}
\eeq
so that
\beqa
{Q}_{ij}^{-1}(s)&=&e^{a_{ij}s}\sqrt{\frac{x_j}{x_i}}B_{ij}(s)\nn
&=& \delta_{ij}+\frac{2\pi\rho
\sqrt{x_ix_j}}{s^3}e^{a_{ij}s}\nn
&&\times\left[L_{ij}^\zero\varphi_2(\sigma_i s)+L_{ij}^\one s\varphi_1(\sigma_i s)\right].
\label{13}
\eeqa
Inserting Eqs.\ \eqref{11} and
\eqref{12} into Eq.\ \eqref{Grfa} we finally get
\beq
G_{ij}(s)=\frac{s^{-2}}{\sqrt{x_i x_j}}\sum_{k=1}^n {P}_{ik}(s)Q_{kj}(s),
\label{15}
\eeq
where use has been made of the additive property \eqref{7}.

We emphasize that Eq.\ \eqref{15} is fully equivalent to Eq.\ \eqref{Grfa} and thus it represents an alternative way of writing the PY solution for AHSs. In both representations the coefficients $L_{ij}^\zero$ and $L_{ij}^\one$ are given by Eq.\ \eqref{LijPY}.
On the other hand, since the structure of Eq.\ \eqref{15} is formally similar to that of the exact solution for 1D NAHSs, Eq.\ \eqref{s1}, it might be expected that Eq.\ \eqref{15} is a reasonable starting point for an extension to 3D NAHSs.

\subsection{Approximation $\aA$}
\label{sec:A}

\subsubsection{The proposal}
A possible proposal for the structural properties of NAHSs is defined by
Eq.\ \eqref{15} with
\beq
{P}_{ij}(s)=\sqrt{x_i x_j}e^{-\sigma_{ij}s}L_{ij}(s),
\label{11NAHS}
\eeq
\beqa
{Q}_{ij}^{-1}(s)&=& \delta_{ij}+\frac{2\pi\rho
\sqrt{x_ix_j}}{s^3}e^{a_{ij}s}\nn
&&\times\left[L_{ij}^\zero\varphi_2(b_{ij} s)+L_{ij}^\one s\varphi_1(b_{ij} s)\right],
\label{13NAHS}
\eeqa
where $L_{ij}(s)$ is still given by Eq.\ \eqref{Lserie} [with $L_{ij}^\zero$ and $L_{ij}^\one$ yet to be determined] and
\beq
b_{ij}\equiv \sigma_{ij}+a_{ij}.
\label{14}
\eeq
Equations \eqref{11NAHS} and \eqref{13NAHS} are obtained from Eqs.\ \eqref{11} and \eqref{13}, respectively, by the extensions $\deltak_{ij}\to \sigma_{ij}$ and $\sigma_i\to b_{ij}$ [compare Eqs.\ \eqref{4} and \eqref{14}].
Note that Eq.\ \eqref{13NAHS} can also be written as
\beq
{Q}_{ij}^{-1}(s)= \delta_{ij}-\frac{2\pi\rho
\sqrt{x_ix_j}}{s^3}\left[N_{ij}(s)e^{a_{ij}s}-L_{ij}(s)e^{-\sigma_{ij}s}\right],
\label{Qij}
\eeq
where
\beq
N_{ij}(s)\equiv L_{ij}^\zero\left(1-b_{ij} s+\frac{b_{ij}^2
  s^2}{2}\right)+L_{ij}^\one s\left(1-b_{ij} s\right).
\label{Nkj}
\eeq

Of course, the coefficients $L_{ij}^\zero$ and $L_{ij}^\one$ are no longer given by Eq.\ \eqref{LijPY} but are obtained from the physical conditions
\beq
\lim_{s\to 0}s^2 G_{ij}(s)=1,
\label{16}
\eeq
\beq
\lim_{s\to 0}s^{-1}\left[s^2 G_{ij}(s)-1\right]=0,
\label{17}
\eeq
which follow from Eq.\ \eqref{s2G}.
To that purpose, it is convenient to rewrite Eq.\ \eqref{15} as
\beq
s^2\sum_{k=1}^n \sqrt{x_i x_k}G_{ik}(s)Q^{-1}_{kj}(s)=P_{ij}(s).
\label{21}
\eeq
Using Eqs.\ \eqref{11NAHS} and \eqref{13NAHS}, Eq.\ \eqref{16} implies
\beq
1+\pi\rho\sum_{k=1}^n x_k b_{kj}^2\left(L_{kj}^\one- \frac{1}{3}L_{kj}^\zero
b_{kj}\right)=L_{ij}^\zero.
\label{22}
\eeq
Likewise, Eq.\ \eqref{17} gives
\beqa
&&\pi\rho\sum_{k=1}^n x_k
b_{kj}^2\left[a_{kj}\left(L_{kj}^\one-\frac{1}{3}L_{kj}^\zero
b_{kj}\right)\right.
\nn
&&
\left.-\frac{1}{3}b_{kj}\left(L_{kj}^\one-\frac{1}{4}L_{kj}^\zero
b_{kj}\right)\right]=L_{ij}^\one-\sigma_{ij}L_{ij}^\zero.
\label{23}
\eeqa
Equations \eqref{22} and \eqref{23} imply that both $L_{ij}^\zero$ and
$L_{ij}^\one-\sigma_{ij}L_{ij}^\zero$ are independent of the subscript
$i$, i.e.,
\beq
L_{ij}^\zero=S_j,\quad L_{ij}^\one=T_j+\sigma_{ij}S_j,
\label{24}
\eeq
where $S_j$ and $T_j$ are determined from Eqs.\ \eqref{22} and
\eqref{23}. The solution is
\beq
S_j=\frac{1-\pi \rho\Psi_j}{\left(1-\pi
\rho\Lambda_j\right)\left(1-\pi
\rho\Psi_j\right)-\pi^2\rho^2\mu_{j|2,0}\Omega_j},
\label{28a}
\eeq
\beq
T_j=\frac{\pi\rho\Omega_j}{\left(1-\pi \rho\Lambda_j\right)\left(1-\pi
\rho\Psi_j\right)-\pi^2\rho^2\mu_{j|2,0}\Omega_j},
\label{29a}
\eeq
where we have called
\beq
\Lambda_j\equiv \mu_{j|2,1}-\frac{1}{3}\mu_{j|3,0},
\label{30a}
\eeq
\beq
\Psi_j\equiv \frac{2}{3}\mu_{j|3,0}-\mu_{j|2,1},
\label{31}
\eeq
\beq
\Omega_j\equiv \mu_{j|3,1}-\mu_{j|2,2}-\frac{1}{4}\mu_{j|4,0},
\label{32a}
\eeq
and
\beq
\mu_{j|p,q}\equiv \sum_{k=1}^n x_k b_{kj}^p \sigma_{kj}^q.
\label{25}
\eeq
In the additive case ($b_{kj}=\sigma_k$) one has
$\Lambda_j=\frac{1}{6}M_3+
\frac{1}{2}M_2\sigma_j$,
$\Psi_j=\frac{1}{6}M_3-
\frac{1}{2}M_2\sigma_j$,
and $\Omega_j=-\frac{1}{4}M_2\sigma_j^2$, so that
$S_j=\theta_1+\theta_2\sigma_j$ and
$T_j=-\frac{1}{2}\theta_2\sigma_j^2$, in agreement with Eq.\ \eqref{LijPY}.
In the case of binary nonadditive mixtures ($\Delta\neq 0$), it can be easily checked that the common denominator in Eqs.\ \eqref{28a} and \eqref{29a} is positive definite. It only vanishes if $\Delta=-2\sigma_2/(\sigma_1+\sigma_2)$ (assuming $\sigma_2\leq \sigma_1$) and $\eta=1+x_2\sigma_2^3/x_1\sigma_1^3$.

Equation \eqref{24} closes the approximation \eqref{15}--\eqref{13NAHS}. It relies on the same philosophy as the so-called rational-function approximation  used in the past for HS and related systems \cite{RS11,HYS08} and, therefore, we will use the acronym RFA to refer to it.
The explicit forms of $G_{ij}(s)$ for binary mixtures ($n=2$) are presented in Appendix \ref{app0}.

\subsubsection{Low-density behavior}
To first order in density, Eqs.\ \eqref{24}--\eqref{29a} yield
\beq
L_{ij}^\zero=1+\pi\rho\Lambda_j+\mathcal{O}(\rho^2),
\eeq
\beq
L_{ij}^\one=\sigma_{ij}+\pi\rho\left(\sigma_{ij}\Lambda_j+\Omega_j\right)+
\mathcal{O}(\rho^2).
\eeq
Thus,
\beqa
{Q}_{ij}(s)&=& \delta_{ij}-\frac{2\pi\rho
\sqrt{x_ix_j}}{s^3}e^{a_{ij}s}\left[\varphi_2(b_{ij} s)+\sigma_{ij} s\varphi_1(b_{ij} s)\right]\nn
&&+
\mathcal{O}(\rho^2).
\label{13rho}
\eeqa
Insertion into Eq.\ \eqref{15} yields
\begin{widetext}
\beqa
G_{ij}(s)&=&\frac{e^{-\sigma_{ij}s}}{s^2}\left(1+\sigma_{ij}s\right)+\pi\rho\frac{e^{-\sigma_{ij}s}}{s^2}\left[\Lambda_j+\left(\sigma_{ij}\Lambda_j+\Omega_j\right)s\right]
-\frac{2\pi\rho}{s^5}\sum_{k=1}^n x_k e^{-(\sigma_{ik}+\sigma_{kj})s}
\nn&&\times
(1+\sigma_{ik}s)(1+\sigma_{kj}s)
+\frac{2\pi\rho}{s^5}\sum_{k=1}^n x_k e^{-(\sigma_{ik}-a_{kj})s}(1+\sigma_{ik}s)
\left[1-a_{kj} s-\frac{1}{2}(\sigma_{kj}^2-a_{kj}^2)s^2\right]\nn
&&+
\mathcal{O}(\rho^2).
\label{Grho2}
\eeqa

Laplace inversion gives
\beqa
g_{ij}(r)&=&\Theta(r-\sigma_{ij})+\frac{\pi\rho}{r}\Theta(r-\sigma_{ij})\left(\Lambda_j r+\Omega_j\right)
-\frac{\pi\rho}{12r}\sum_{k=1}^n x_k \Theta(r-\sigma_{ik}-\sigma_{kj})(r-\sigma_{ik}-\sigma_{kj})^2\nn
&&\times\left[r^2+2(\sigma_{ik}+\sigma_{kj})r-3(\sigma_{ik}-\sigma_{kj})^2\right]+\frac{\pi\rho}{12r}\sum_{k=1}^n x_k \Theta(r-\sigma_{ik}+a_{kj})(r-\sigma_{ik}+a_{kj})\nn
&&\times\left[r^3+(\sigma_{ik}-a_{kj})r^2-(5\sigma_{ik}^2+6\sigma_{kj}^2+2\sigma_{ik}a_{kj}-a_{kj}^2)r\right.\nn
&&\left.+3(\sigma_{ik}+a_{kj})(\sigma_{ik}^2+a_{kj}^2-2\sigma_{kj}^2)\right]
+
\mathcal{O}(\rho^2).
\label{grho2}
\eeqa
\end{widetext}

As a consequence, approximation {$\aA$} is consistent with the exact limits \eqref{cond_iii} and \eqref{Grho}.
To first order in density,  the approximation
correctly accounts for singularities of $g_{ij}(r)$ at distances $r=\sigma_{ij}$ and
$r=\sigma_{ik}+\sigma_{kj}$, $k=1,\ldots,n$ [see Eq.\ \eqref{grho2exact}]. On the other hand, {we see from Eq.\ \eqref{Grho2} that, already  to first order in density,} approximation $\aA$  introduces spurious
singularities  at  $r=\sigma_{ik}-a_{kj}\neq \sigma_{ij}$.
One might even have
$d_{ij;k}\equiv\sigma_{ik}-a_{kj}- \sigma_{ij}<0$. In particular,
$d_{ii;k}=\deltak_{ik}\Delta_{ik}$ becomes negative
if $\Delta_{ik}<0$. Analogously,
$d_{ij;i}=-\deltak_{ij}\Delta_{ij}$ is negative if
$\Delta_{ij}>0$.
Therefore, approximation {$\aA$}
does not verify in general the condition \eqref{cond_i}.
It is worth noting, however, that the hard-core condition \eqref{cond_i} is also typically violated by density-functional theories \cite{S07b}.
The inability of approximation $\aA$ to guarantee that $g_{ij}(r)=0$ for $r<\sigma_{ij}$ will be remedied by approximation $\aB$ described in Sec.\ \ref{sec:A+}.

\subsubsection{Short-range behavior}
\begin{figure}
\includegraphics[width=.9\columnwidth]{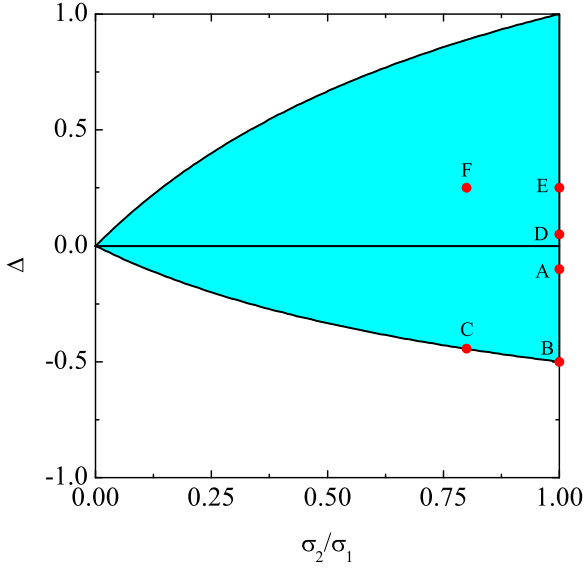}
\caption{(Color online) Plane $\Delta$ vs $\sigma_2/\sigma_1$ showing the
  shaded region $-\sigma_2/(\sigma_1+\sigma_2)\leq\Delta\leq
  2\sigma_2/(\sigma_1+\sigma_2)$ where   the first two singularities of $g_{ij}(r)$, according to approximation $\aA$, are $\sigma_{ij}$ and $\sigma_{ik}-a_{kj}$ with $k\neq j$. The circles denote the systems
  analyzed in Sec.\ \protect\ref{sec:structure}.\label{diagram2}}
\end{figure}

Before presenting approximation $\aB$, we will need to restrict ourselves to cases where the first two singularities of $g_{ij}(r)$, as given by approximation $\aA$, are $\sigma_{ij}$ and $\s_{ij}\equiv\min(\sigma_{ik}-a_{kj};k=1,\ldots,n; k\neq j)$. As proven in
Appendix \ref{app1}, the above requirement in the binary case ($n=2$) implies the constraint $-\sigma_2/(\sigma_1+\sigma_2)\leq\Delta\leq
2\sigma_2/(\sigma_1+\sigma_2)$, where, without loss of generality, it has been assumed $\sigma_2\leq\sigma_1$.
This region of applicability is shown in Fig.\
\ref{diagram2}.

Appendix \ref{app2} gives the expressions for $g_{ij}(r)$ in the range $0\leq r\leq \max(\sigma_{ij},\s_{ij})+\epsilon$, where $\epsilon$ is any positive value smaller than the separation between $\max(\sigma_{ij},\s_{ij})$ and the next singularity of $g_{ij}(r)$, provided by approximation {$\aA$} for binary mixtures. Extending to general $n$ the arguments presented there, we can write
\beq
G_{ij}(s)=e^{-\sigma_{ij}s}\Phi_{ij}(s)+2\pi\rho x_\kappa e^{-\s_{ij}s}\Gamma_{i\kappa j}(s)+\cdots,
\label{Gijshort}
\eeq
where   $k=\kappa$ is the index corresponding to $\s_{ij}$, i.e., $\s_{ij}=\sigma_{i\kappa}-a_{\kappa j}$, and the ellipsis denotes terms headed by exponentials of the form $e^{-\lambda s}$ with $\lambda > \max(\sigma_{ij},\s_{ij})$. In Eq.\ \eqref{Gijshort},
\beq
\Phi_{ij}(s)\equiv \frac{1}{s^2}L_{ij}(s)\bar{Q}_{jj}(s),
\label{Phiij}
\eeq
\beq
\Gamma_{ikj}(s)\equiv\frac{1}{s^5} \frac{L_{ik}(s)N_{kj}(s)}{D_0(s)},
\label{55}
\eeq
where
\beq
\bar{Q}_{ij}^{-1}(s)\equiv \delta_{ij}-\frac{2\pi\rho
\sqrt{x_ix_j}}{s^3}N_{ij}(s),
\label{56}
\eeq
and $D_0(s)$ is the determinant of the matrix $\bar{\mathsf{Q}}^{-1}(s)$.
Explicit expressions of $\Phi_{ij}(s)$ and $D_0(s)$ for binary mixtures are given in Appendix \ref{app2}.

Taking the Laplace inversion of Eq.\ \eqref{Gijshort}, one finds that,  in the interval $0\leq r\leq
\max(\sigma_{ij},\s_{ij})+\epsilon$,
\beqa
g_{ij}(r)&=&\frac{1}{r}\Theta(r-\sigma_{ij})\phi_{ij}
(r-\sigma_{ij})
  \nn &&+\frac{2\pi\rho}{r}
  x_\kappa
   \Theta(r-\s_{ij})
\gamma_{i\kappa j}(r-\s_{ij}),
\label{gijshort}
\eeqa
where  $\phi_{ij}(r)$ and $\gamma_{ikj}(r)$ are the inverse Laplace transforms of
$\Phi_{ij}(s)$ and $\Gamma_{ikj}(s)$, respectively.

Note that
$\phi_{ij}(0)=\lim_{s\to\infty}\Phi_{ij}(s)=L_{ij}^\one$, while
$\gamma_{ikj}(0)=\lim_{s\to\infty}\Gamma_{ikj}(s)=0$.
Therefore, the contact values  are
\beq
g_{ij}(\sigma_{ij}^+)=\frac{L_{ij}^\one}{\sigma_{ij}}+
\frac{2\pi\rho}{\sigma_{ij}}
x_\kappa\Theta(\sigma_{ij}-\s_{ij})\gamma_{i\kappa j}
(\sigma_{ij}-\s_{ij}).
\label{85}
\eeq
As expected, Eq.\ \eqref{85} reduces to Eq.\ \eqref{contact-value} in the additive case.

\subsection{Approximation $\aB$}
\label{sec:A+}
This new option for $g_{ij}(r)$ will differ from approximation {$\aA$} only in the region
$\min(\sigma_{ij},\s_{ij})\leq r\leq \max(\sigma_{ij},\s_{ij})$. More specifically,
\beqa
\left.g_{ij}(r)\right|_{\aB}&=&\left.g_{ij}(r)\right|_{\aA}
+\frac{2\pi\rho}{r}
x_\kappa\left[\Theta(r-\sigma_{ij})\right.\nn
&&\left.-\Theta(r-\s_{ij})\right]
\gamma_{i\kappa j}(r-\s_{ij}).
\label{52}
\eeqa
On account of Eq.\ \eqref{gijshort}, Eq.\ \eqref{52} can be equivalently rewritten as
\beq
\left.g_{ij}(r)\right|_{\aB}=\begin{cases}
\Theta(r-\sigma_{ij})\left.g_{ij}(r)\right|_{\aA},\quad\s_{ij}<\sigma_{ij},\\
\left.g_{ij}(r)\right|_{\aA}+\Theta(r-\sigma_{ij})\Theta(\s_{ij}-r)\\
\quad\times \frac{2\pi\rho}{r}
x_\kappa\gamma_{i\kappa j}(r-\s_{ij}),\quad\s_{ij}>\sigma_{ij}.
\end{cases}
\label{52bis}
\eeq
We see from Eq.\ \eqref{52bis} that the idea behind approximation {$\aB$} is two-fold. On the one hand, it removes the unphysical violation of the property $g_{ij}(r)=0$ for
$r<\sigma_{ij}$ that is present in option {$\aA$} when $\s_{ij}<\sigma_{ij}$. On the other hand, if
$\s_{ij}>\sigma_{ij}$, approximation {$\aB$}
extrapolates to the region $\sigma_{ij}<r<\s_{ij}$
 the functional form of $g_{ij}(r)$ provided by approximation {$\aA$} in the region between
$\s_{ij}$ and the next singularity.

In the interval $0\leq r\leq \max(\sigma_{ij},\s_{ij})+\epsilon$,
\beq
\left.g_{ij}(r)\right|_{\aB}=\frac{1}{r}\Theta(r-\sigma_{ij})[\phi_{ij}
(r-\sigma_{ij})
    +{2\pi\rho} x_\kappa
\gamma_{i\kappa j}(r-\s_{ij})].
\label{gijshort2}
\eeq
In particular,
\beq
\left.g_{ij}(\sigma_{ij}^+)\right|_{\aB}=\frac{L_{ij}^\one}{\sigma_{ij}}+
\frac{2\pi\rho}{\sigma_{ij}}
x_\kappa\gamma_{i\kappa j}
(\sigma_{ij}-\s_{ij}).
\label{85b}
\eeq

\subsection{Approximation $\aB^{(m)}$}
In approximation $\aB$ the
\emph{full} functional form of $\gamma_{ikj}(r)$ is used. This can  create some artificial problems in the region
$\sigma_{ij}<r<\s_{ij}$ when $\s_{ij}>\sigma_{ij}$ and the distance
$\s_{ij}-\sigma_{ij}$ is rather large (as happens in the WR model). Reciprocally, if $\s_{ij}-\sigma_{ij}$ is not large, it becomes unnecessarily complicated  to consider the entire nonlinear function $\gamma_{ikj}(r)$ in the interval $\sigma_{ij}<r<\s_{ij}$.
Thus,  we now propose a variant of approximation $\aB$, here denoted as $\aB^{(m)}$, whereby the full true
function $\gamma_{i\kappa j}(r)$ is preserved if $\s_{ij}<\sigma_{ij}$ (in order to
enforce the physical constraint of a vanishing RDF for
$r<\sigma_{ij}$) but is replaced
by its $m$th degree polynomial approximation  $\gamma_{i\kappa j}^{(m)}(r)$ if $\s_{ij}>\sigma_{ij}$.
In summary, option $\aB^{(m)}$ is defined by
\beq
\left.g_{ij}(r)\right|_{\aB^{(m)}}=\begin{cases}
\Theta(r-\sigma_{ij})\left.g_{ij}(r)\right|_{\aA},\quad\s_{ij}<\sigma_{ij},\\
\left.g_{ij}(r)\right|_{\aA}+\Theta(r-\sigma_{ij})\Theta(\s_{ij}-r)\\
\quad\times \frac{2\pi\rho}{r}
x_\kappa\gamma_{i\kappa j}^{(m)}(r-\s_{ij}),\quad\s_{ij}>\sigma_{ij}.
\end{cases}
\label{54}
\eeq
Consequently, the contact values are
\beqa
\left.g_{ij}(\sigma_{ij}^+)\right|_{\aB^{(m)}}&=&\frac{L_{ij}^\one}{\sigma_{ij}}+
\frac{2\pi\rho}{\sigma_{ij}}
x_\kappa\nn
&&\times \Big[\Theta(\sigma_{ij}-\s_{ij})\gamma_{i\kappa j}
(\sigma_{ij}-\s_{ij})\nn
&&+\Theta(\s_{ij}-\sigma_{ij})\gamma_{i\kappa j}^{(m)}
(\sigma_{ij}-\s_{ij})\Big].
\label{85b2}
\eeqa

The polynomial $\gamma_{ikj}^{(m)}(r)$ is obtained by truncating after $r^{m}$ the expansion of  $\gamma_{ikj}(r)$ in powers of $r$. Such an expansion is directly related to that of the Laplace transform $\Gamma_{ikj}(s)$ in powers of $s^{-1}$.
For large $s$, $\Gamma_{ikj}(s)$ can be shown to be given by
\begin{widetext}
  \beqa
\Gamma_{ikj}(s)&=&s^{-2}L_{ik}^\one\left[L_{kj}^\zero\frac{b_{kj}}{2}-
L_{kj}^\one\right]b_{kj}+
s^{-3}\Bigg\{L_{ik}^\zero\left[L_{kj}^\zero\frac{b_{kj}}{2}-
L_{kj}^\one\right]b_{kj}-
L_{ik}^\one\left[L_{kj}^\zero{b_{kj}}-L_{kj}^\one\right]\nn
&&+
2\pi\rho L_{ik}^\one\left[L_{kj}^\zero\frac{b_{kj}}{2}-
L_{kj}^\one\right]b_{kj}\sum_{\ell=1}^n x_\ell
\left[L_{\ell\ell}^\zero\frac{\sigma_{\ell}}{2}-
L_{\ell\ell}^\one\right]\sigma_{\ell}\Bigg\}+\mathcal{O}(s^{-4}).
\label{58}
\eeqa
Consequently, the linear and quadratic approximations are
\beq
\gamma_{ikj}^{(1)}(r)=
L_{ik}^\one\left[L_{kj}^\zero\frac{b_{kj}}{2}-
L_{kj}^\one\right]b_{kj}{r},
\label{gamma1}
\eeq
\beqa
\gamma_{ikj}^{(2)}(r)&=&\gamma_{ikj}^{(1)}(r)+
\Bigg\{L_{ik}^\zero\left[L_{kj}^\zero\frac{b_{kj}}{2}-
L_{kj}^\one\right]b_{kj}-
L_{ik}^\one\left[L_{kj}^\zero{b_{kj}}-L_{kj}^\one\right]\nn
&&+
2\pi\rho L_{ik}^\one\left[L_{kj}^\zero\frac{b_{kj}}{2}-
L_{kj}^\one\right]b_{kj}\sum_{\ell=1}^n x_\ell
\left[L_{\ell\ell}^\zero\frac{\sigma_{\ell}}{2}-
L_{\ell\ell}^\one\right]\sigma_{\ell}\Bigg\}\frac{r^2}{2}.
\label{gamma2}
\eeqa
\end{widetext}

 Of course, the three sets of approximations $\aA$, $\aB$, and $\aB^{(m)}$ reduce to the PY solution in the additive case.
 Obviously, $\aB\equiv\aB^{(\infty)}$.
In Sec.\ \ref{sec:structure} we will generally use $\aB^{(1)}$.

\section{Comparison with Monte Carlo simulations for binary mixtures. The equation of state}
\label{sec:EOS}

The compressibility factor $Z$ is obtained via the virial and compressibility routes by Eqs.\ (\ref{Zv}) and (\ref{Zc}), respectively.
In the case of the virial route one needs  the
contact values of the RDF, which are given by Eqs.\ (\ref{85}), \eqref{85b}, and \eqref{85b2} in approximations
{$\aA$}, $\aB$, and $\aB^{(m)}$, respectively.

In the compressibility route, the isothermal compressibility $\chi$ is obtained from Eq.\ \eqref{chih}, where $\widehat{h}_{ij}(0)=\rho\sqrt{x_ix_j}\widetilde{h}_{ij}(0)=-4\pi \rho\sqrt{x_ix_j}H_{ij}^\one$, $H_{ij}^\one$ being the coefficient of $s^3$ in the series expansion of $s^2 G_{ij}(s)$ in powers of $s$ [cf.\ Eq.\ \eqref{s2G}]. We recall that $G_{ij}(s)$ is given by Eq.\ \eqref{15} in approximation $\aA$. In approximations $\aB$ and  $\aB^{(m)}$, Eqs.\ \eqref{52} and \eqref{54} imply that
\beq
\left.H_{ij}^\one\right|_{\aB}=\left.H_{ij}^\one\right|_{\aA}-2\pi\rho x_\kappa\int_{\sigma_{ij}}^{\s_{ij}}dr\, r\gamma_{i\kappa j}(r-\s_{ij}),
\label{Hij}
\eeq
\beqa
\left.H_{ij}^\one\right|_{\aB^{(m)}}&=&\left.H_{ij}^\one\right|_{\aA}-2\pi\rho x_\kappa\int_{\sigma_{ij}}^{\s_{ij}}dr\, r\Big[\Theta(\sigma_{ij}-\s_{ij})\nn
&&\times\gamma_{i\kappa j}(r-\s_{ij})+\Theta(\s_{ij}-\sigma_{ij})\gamma_{i\kappa j}^{(m)}(r-\s_{ij})\Big].\nn
\label{Hijm}
\eeqa

In any case, for the sake of simplicity, we will restrict ourselves in most of this section to approximation $\aA$.

\subsection{Dependence of the EOS on nonadditivity}

Here we study the dependence of the EOS on the nonadditivity parameter $\Delta$ by fixing
all the other parameters of the mixture (density, composition, and size ratio).

\subsubsection{Symmetric binary mixtures}

Symmetric mixture are obtained when
$\sigma_1=\sigma_2=\sigma$. Therefore, in the additive case ($\Delta=0$) one recovers the one-component HS system, i.e., $g_{11}(r)=g_{22}(r)=g_{12}(r)=g(r)$, regardless of the value of $x_1$.
\begin{figure}
\begin{center}
\includegraphics[width=0.9\columnwidth]{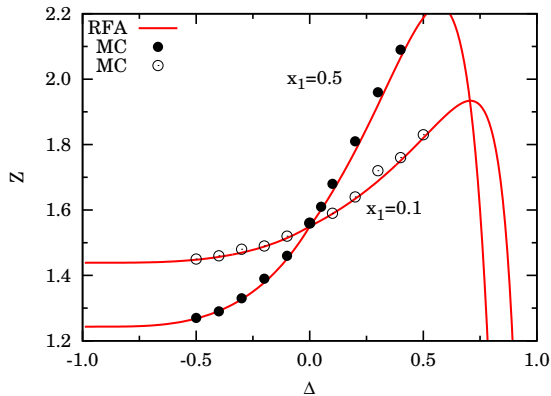}
\end{center}
\caption{(Color online) Compressibility factor as a function of the nonadditivity
parameter for a symmetric binary mixture of NAHSs
at $\rho\sigma^3=0.2$ and two different compositions. The MC data are taken from
Refs.\ \protect\cite{JJR94a,JJR94b}.}
\label{fig:Zs}
\end{figure}

Figure  \ref{fig:Zs} compares the compressibility factor obtained from MC simulations \cite{JJR94a,JJR94b} with that predicted by approximation $\aA$ for some representative symmetric systems. We observe that approximation {$\aA$} reproduces quite well the
exact simulation data at all values of the nonadditivity parameter. At this  low density ($\rho\sigma^3=0.2, \eta\simeq 0.105$) the virial and compressibility
routes are not distinguishable on the scale of the graph.

\subsubsection{Asymmetric binary mixtures}

Asymmetric mixtures correspond to
$\sigma_1\neq\sigma_2$. In that case,  when $\Delta=0$ one
recovers the AHS mixture.
\begin{figure}
\begin{center}
\includegraphics[width=0.9\columnwidth]{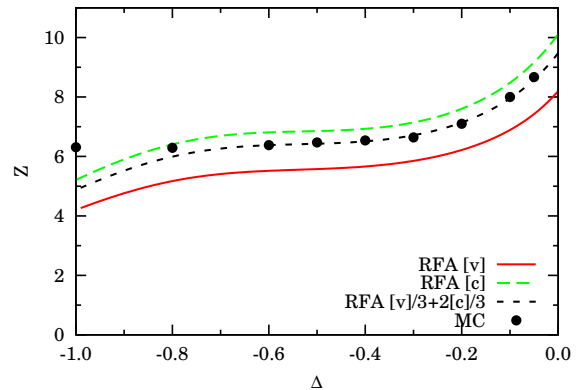}
\end{center}
\caption{(Color online) Compressibility factor as a function of the nonadditivity
parameter for an equimolar asymmetric binary mixture of NAHSs with a size ratio $\sigma_2/\sigma_1=1/3$ at a packing fraction
$\eta=0.5$. The symbols [v] and [c] stand for the virial and compressibility routes, respectively. The MC data are taken from Ref. \protect\cite{H97}.
\label{fig:Za}}
\end{figure}

Figure \ref{fig:Za} shows the $\Delta$ dependence of $Z$ for negative nonadditivity and an equimolar ($x_1=x_2=\frac{1}{2}$) asymmetric mixture ($\sigma_2/\sigma_1=1/3$) at a relatively large density ($\eta=0.5$). In this case the virial route of approximation $\aA$ underestimates the values of $Z$, while the compressibility route overestimates them. This is also a typical behavior of the PY equation for AHSs. It is thus tempting to try the $Z=\frac{1}{3}Z^v+\frac{2}{3}Z^c$ interpolation recipe \cite{B70,MCSL71,GH72,LL73}, which is known to work well in the additive case.
From Fig.\ \ref{fig:Za} we see that indeed the interpolation formula, as applied to
approximation {$\aA$},  reproduces quite well the exact simulation data, except for $\Delta
\lesssim -0.8$.

\subsection{Dependence of the EOS on the size ratio}

Next,  we study the dependence of Z on the size ratio $\sigma_2/\sigma_1$ by fixing
all the other parameters of the mixture (density, composition, and nonadditivity).

\begin{figure}
\begin{center}
\includegraphics[width=0.9\columnwidth]{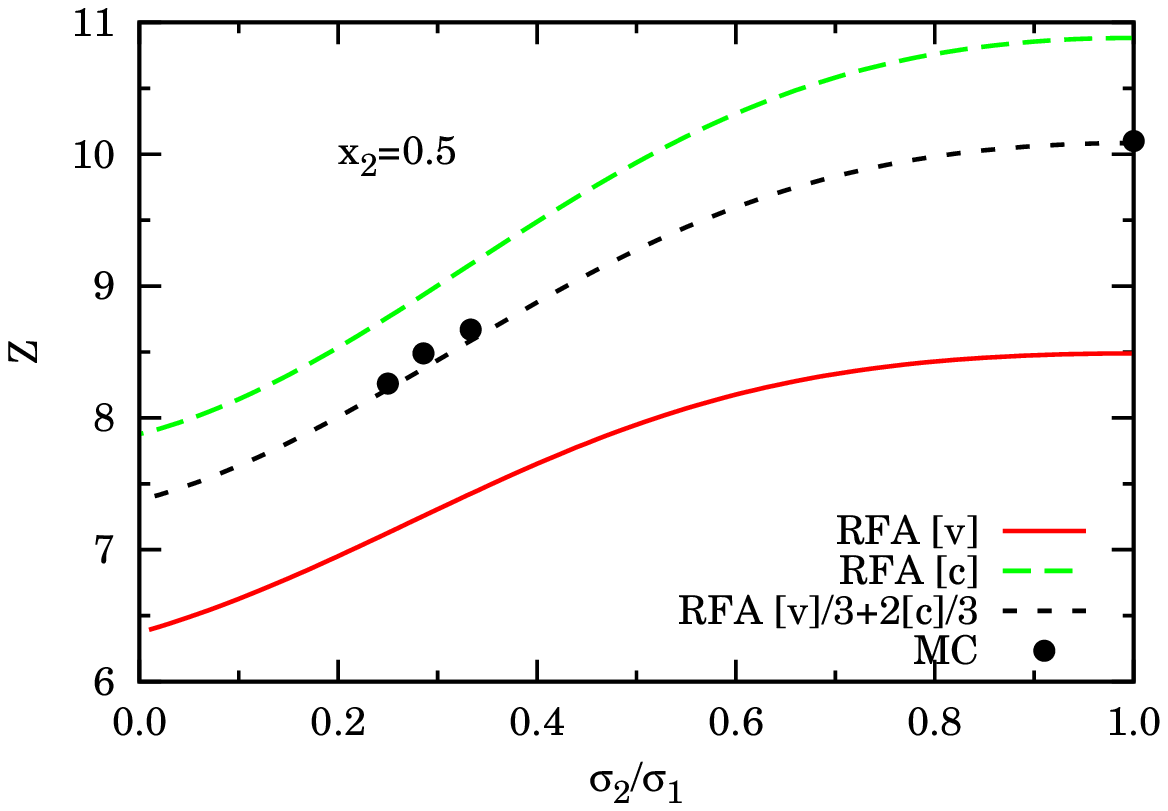}\\
\includegraphics[width=0.9\columnwidth]{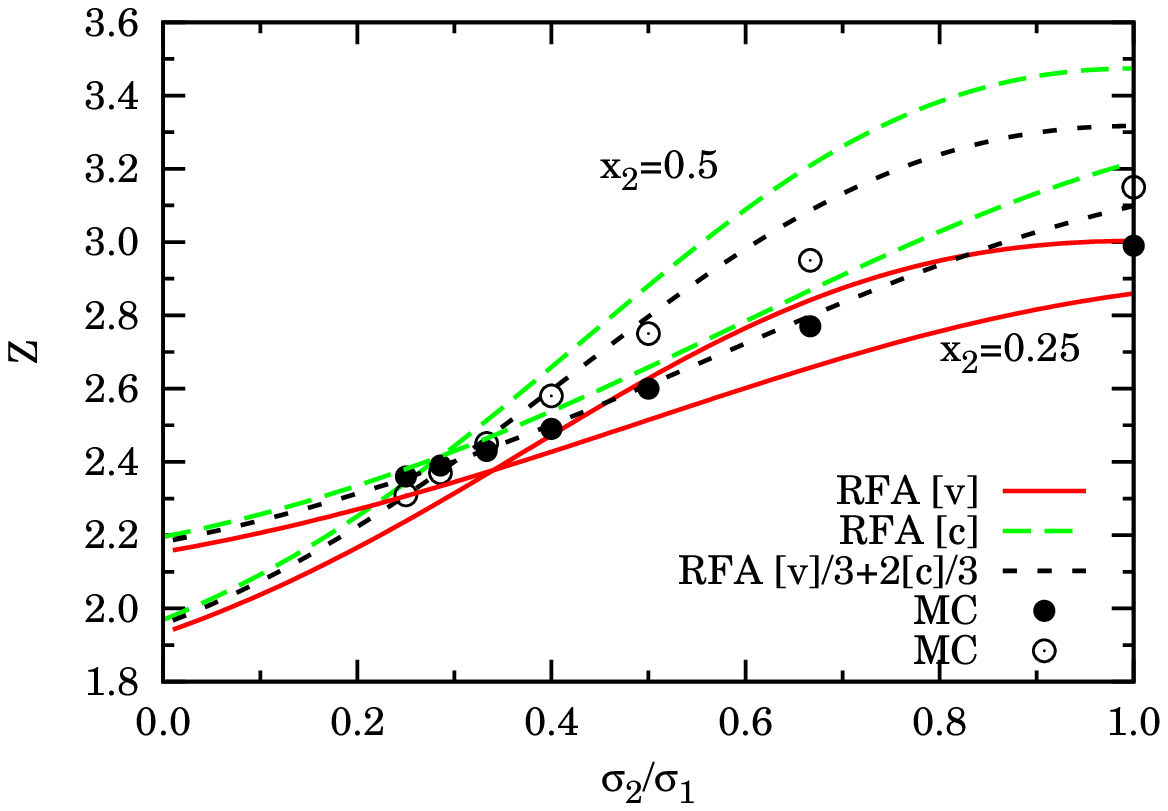}\\
\includegraphics[width=0.9\columnwidth]{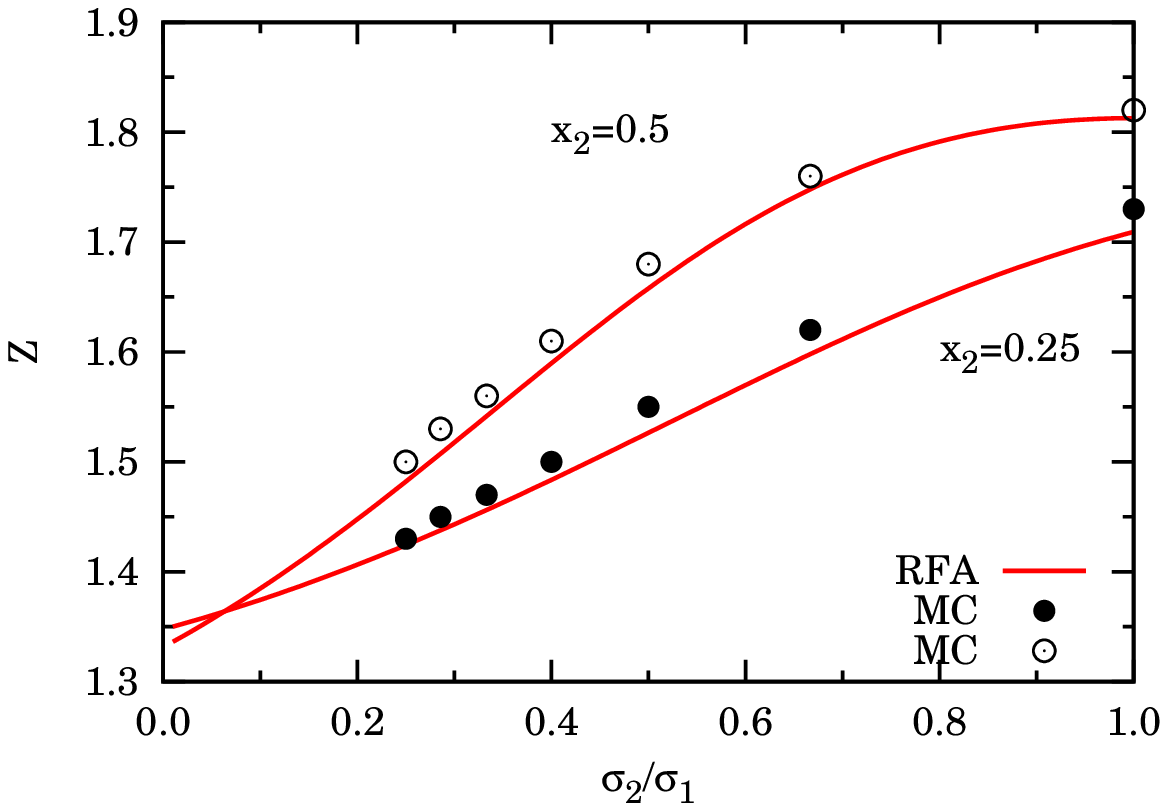}
\end{center}
\caption{(Color online) Compressibility factor as a function of the size ratio
$\sigma_2/\sigma_1$ for binary asymmetric NAHS mixtures  with $x_2=\frac{1}{2}$, $\Delta=-0.05$, and $\eta=0.5$ (top
panel); $x_2=\frac{1}{4},\frac{1}{2}$, $\Delta=0.2$, and $\eta=0.2$ (middle panel);
$x_2=\frac{1}{4},\frac{1}{2}$, $\Delta=0.5$, and $\eta=0.075$ (bottom panel).   In the bottom panel only the theoretical data obtained from the virial
route are shown since they  practically coincide with those obtained from the compressibility route. The MC data are taken from
Ref.\ \cite{H97}.}
\label{fig:Zar}
\end{figure}

The three panels of Fig.\ \ref{fig:Zar} show $Z$ vs $\sigma_2/\sigma_1$ for a slightly negative nonadditivity $\Delta=-0.05$ (top panel),
a moderate positive nonadditivity $\Delta=0.2$ (middle panel), and a larger positive nonadditivity $\Delta=0.5$ (bottom panel).
We observe again that the interpolation recipe $Z=\frac{1}{3}Z^v+\frac{2}{3}Z^c$  for
approximation {$\aA$} agrees well with the exact simulation data, with the
exception of a region close to the size symmetric mixture ($\sigma_2/\sigma_1=1$) for positive nonadditivity and moderate density (middle panel).

\subsection{Contact values}

In Sec.\ \ref{sec:structure} we will analyze the RDF $g_{ij}(r)$ predicted by approximations $\aA$ and $\aB^\one$. Before doing so, and as a bridge between the thermodynamic and structural properties, it is worth considering the contact values.
Table \ref{tab:contact} provides the contact values for some binary equimolar symmetric NAHS mixtures
($\sigma_1=\sigma_2=\sigma$, $x_1=x_2=\frac{1}{2}$), as obtained from MC simulations \cite{BPGG86}, numerical solutions of the PY integral equation \cite{BPGG86}, and our approximations $\aA$ [Eq.\  (\ref{85})] and $\aB^\one$ [Eq.\  (\ref{85b2})].
Since for binary symmetric mixtures $\s_{11}=\s_{22}=\sigma_{12}=\sigma(1+\Delta)$ and $\s_{12}=\sigma$, it turns out that $g_{11}(\sigma^+)=g_{22}(\sigma^+)$ is common in approximations $\aA$ and $\aB^\one$ if $\Delta<0$, while $g_{12}(\sigma_{12}^+)$ is common in both approximations if $\Delta>0$.

\begin{table}
\caption{Contact values for some binary equimolar symmetric NAHS mixtures. The MC and PY data were
taken from Ref.\ \protect\cite{BPGG86}. The labels correspond to systems common to those listed in Table \protect\ref{tab:cases}.
\label{tab:contact}}
\begin{ruledtabular}
\begin{tabular} {cdccccc}
Label&  \Delta &$\rho\sigma^3$ &  Source &
 $g_{11}(\sigma^+)$ & $g_{12}(\sigma_{12}^+)$\\
\hline
  D& 0.05 &$0.8$ & MC             & $5.305$ & $3.762$ \\
   & &      & PY             & $4.451$ & $3.516$ \\
   & &      & $\aA$              & $4.006$ & $3.617$ \\
      &  &      & $\aB^\one$       & $4.580$ & $3.617$ \\
    &&&&&\\
& 0.0  &$0.8$ & MC             &$ 3.971$ & $3.971$ \\
   & &      & PY             & $3.581$ & $3.581 $\\
   & &      & $\aA$              & $3.581$ & $3.581$ \\
   & &      & $\aB^\one$       & $3.581$ & $3.581$ \\
  &  &&&&\\
& -0.05&$0.8 $& MC             & $3.117$ & $3.801$ \\
  &  &      & PY             & $2.925$ & $3.394$ \\
  &  &      & $\aA$              & $2.971$ & $3.148 $\\
   &     &      &$\aB^\one$       & $2.971$ & $3.445$ \\
    &&&&&\\
A& -0.1 &$1.0$ & MC             & $3.394 $& $5.363 $\\
  &  &      & PY             & $3.209$ &$ 4.395$ \\
   & &      & $\aA$             & $3.497$ & $3.883$ \\
    &    &      & $\aB^\one$       & $3.497$ & $4.763$ \\
 &&&&&\\
& -0.3 &$1.0$ & MC             &$ 2.168 $& $2.798$ \\
&    &      & PY             &$ 2.141$ & $2.543$ \\
&    &      & $\aA$              & $2.441$ & $2.251$ \\
&        &      & $\aB^\one$       & $2.441$ & $2.875$ \\
& &&&&\\
B& -0.5 & $1.0$ &MC             & $2.103$ & $1.528$ \\
  &  &      & PY             & $2.060$ & $1.493$ \\
   & &      & $\aA$             & $2.139$ & $1.407 $\\
    &    &      & $\aB^\one$       & $2.139$ & $1.279$ \\
\end{tabular}
\end{ruledtabular}
\end{table}

{}From Table \ref{tab:contact} we observe that   approximation $\aB^\one$
is  superior to the PY theory in estimating the true contact
values, both for positive and negative nonadditivity, except in the cases of $g_{11}(\sigma^+)$ for $\rho\sigma^3=1$ and $\Delta=-0.3$ and of $g_{12}(\sigma_{12}^+)$ for $\rho\sigma^3=1$ and $\Delta=-0.5$.

\section{Comparison with Monte Carlo simulations for binary mixtures. The  structure}
\label{sec:structure}
\begin{table}
\caption{The six binary NAHS mixtures  considered in the analysis of the structure. The last column gives the compressibility factor as obtained from our MC simulations.
\label{tab:cases}}
\begin{ruledtabular}
\begin{tabular}{ccdcccc}
Label &  $\sigma_2/\sigma_1$ &\Delta & $x_1$
&  $\rho\sigma_1^3$ &  $\eta$  & $Z_{\text{MC}}$ \\
\hline
A &$ 1    $& -0.1& $1/2$& $1.0$& $0.5236$& $8.648$ \\
B & $1$& -0.5& $1/2$& $1.0$& $0.5236$& $3.429$ \\
C& $4/5$&-0.444& $1/3$& $1.0$& $0.3533$& $2.335$ \\
D &$ 1$& 0.05& $1/2$& $0.8$& $0.4189$& $9.083$ \\
E &$ 1$& 0.25& $1/2$& $0.3$ &$0.1571 $& $2.556$ \\
F &$ 4/5 $&0.25& $1/3$&$0.3$&$ 0.1060$&$ 1.876$ \\
\end{tabular}
\end{ruledtabular}
\end{table}
\begin{figure}
\begin{center}
\includegraphics[width=0.9\columnwidth]{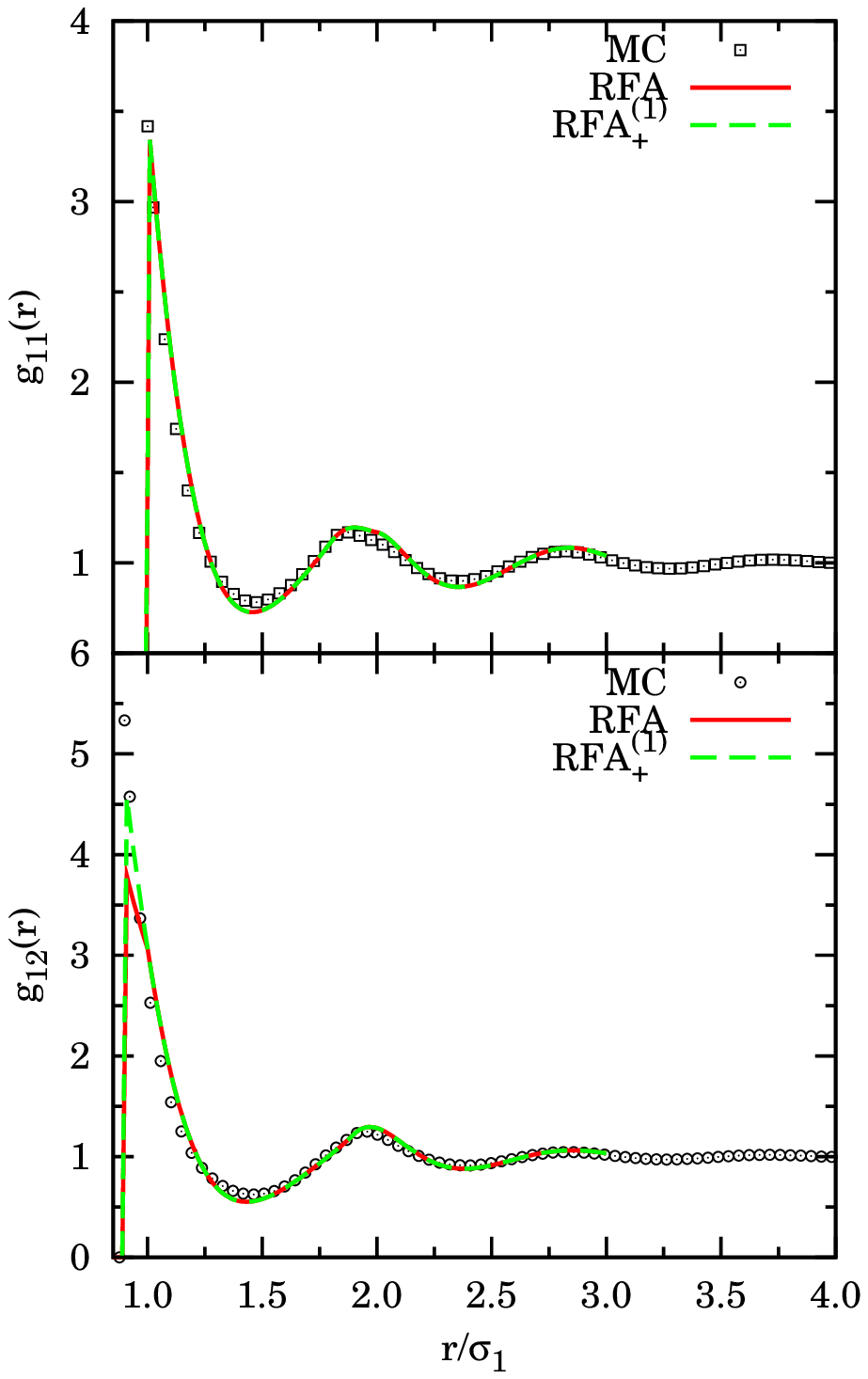}
\end{center}
\caption{(Color online) RDF for system A of table \protect\ref{tab:cases}.
\label{fig:gr-A}}
\end{figure}
\begin{figure}
\begin{center}
\includegraphics[width=0.9\columnwidth]{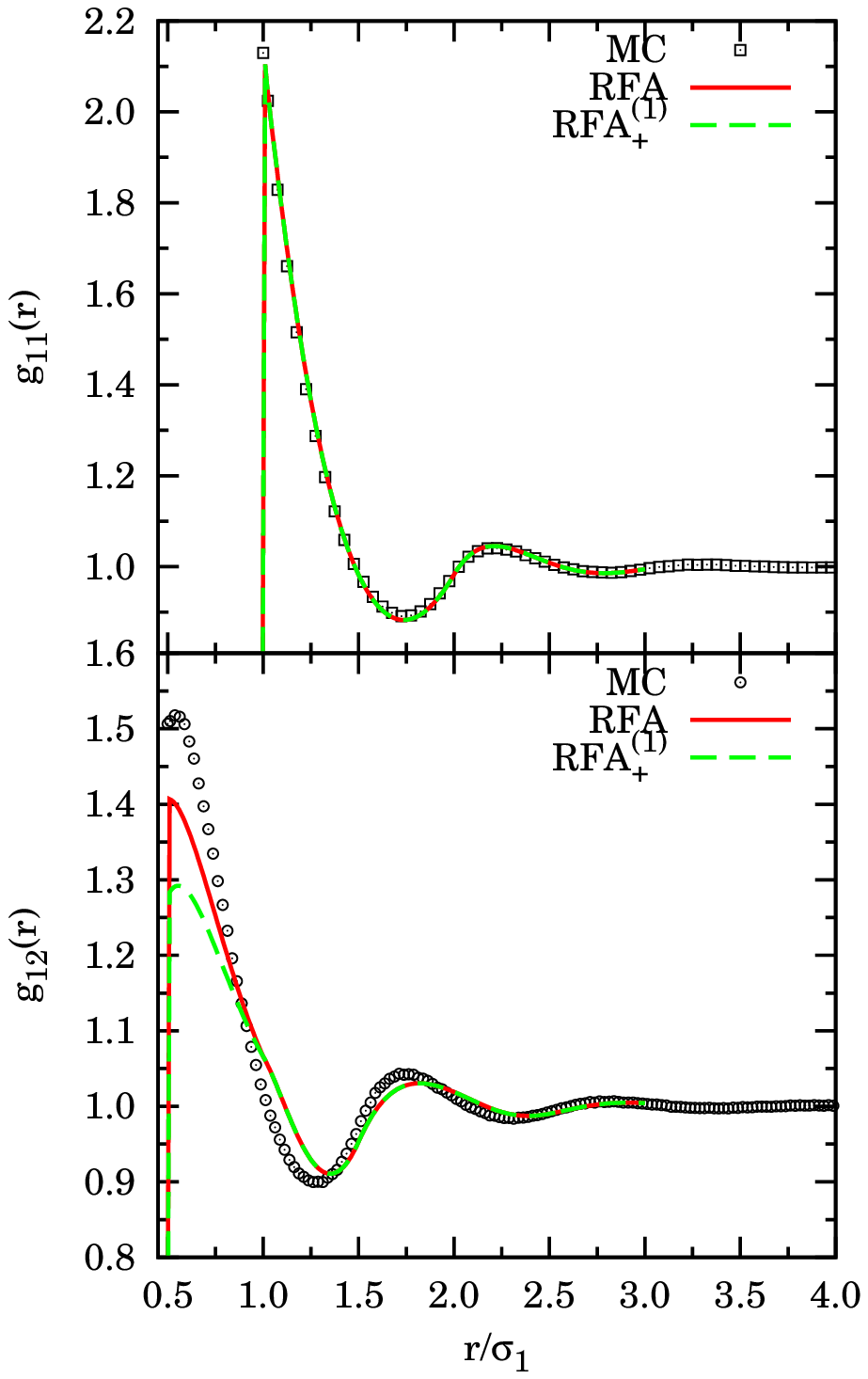}
\end{center}
\caption{(Color online) RDF for system B of table \protect\ref{tab:cases}.
\label{fig:gr-B}}
\end{figure}
\begin{figure}
\begin{center}
\includegraphics[width=0.9\columnwidth]{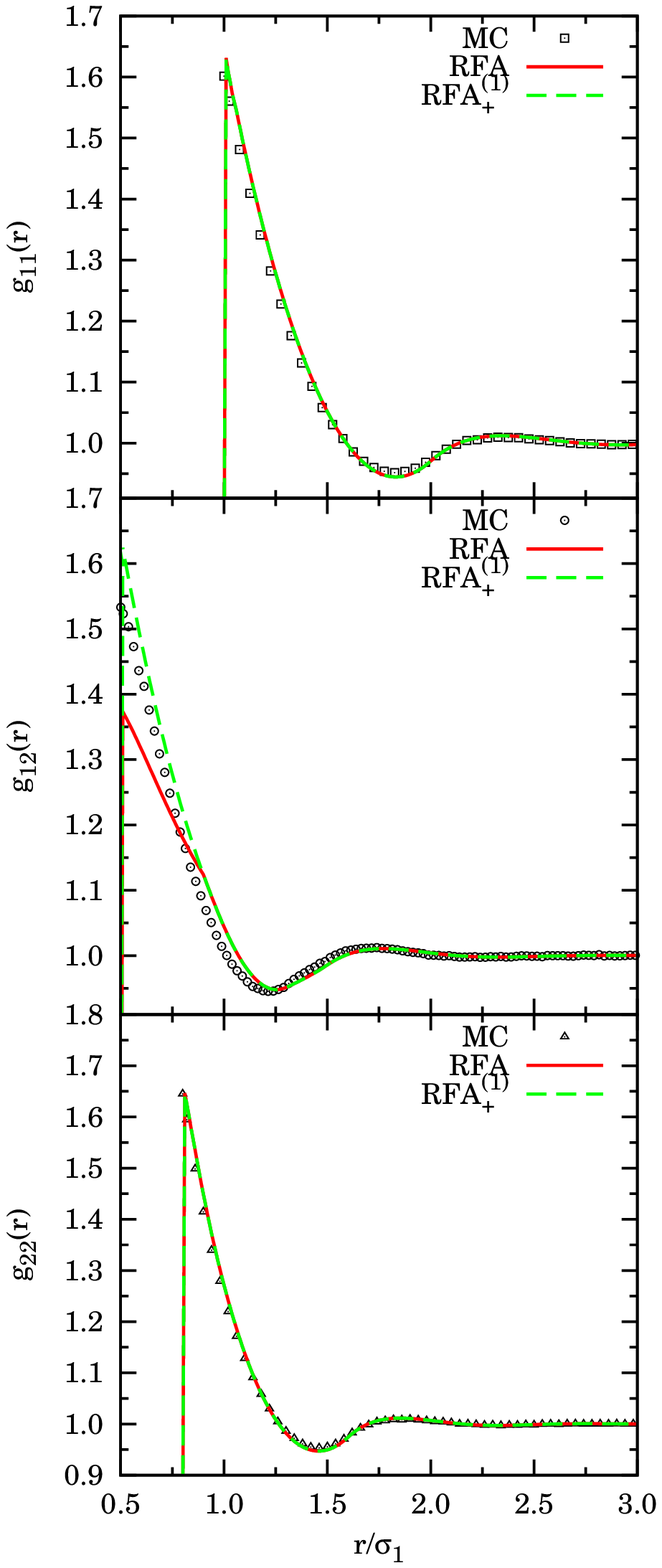}
\end{center}
\caption{(Color online) RDF for system C of table \protect\ref{tab:cases}.}
\label{fig:gr-C}
\end{figure}
\begin{figure}
\begin{center}
\includegraphics[width=0.9\columnwidth]{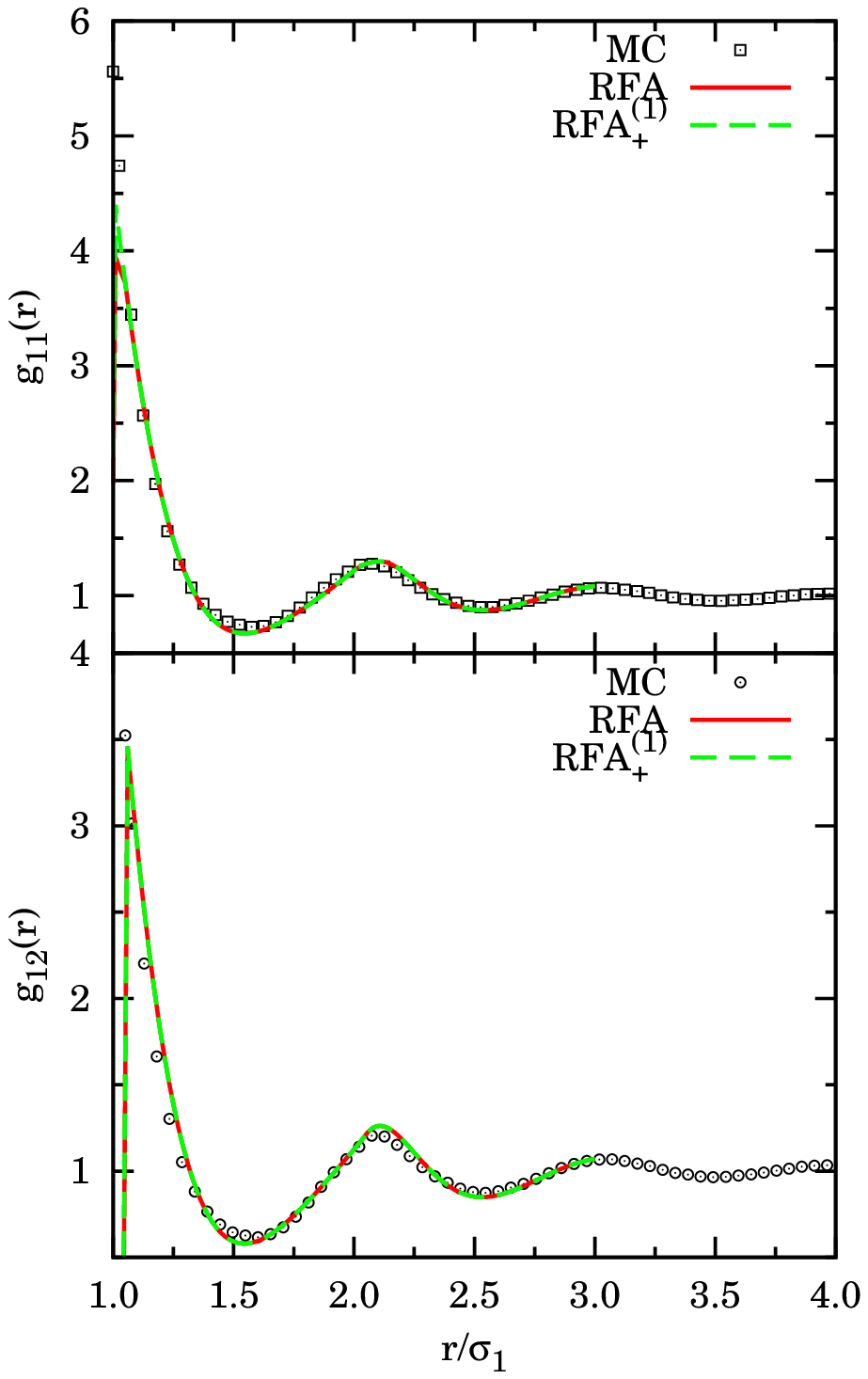}
\end{center}
\caption{(Color online) RDF for system D of table \protect\ref{tab:cases}.}
\label{fig:gr-D}
\end{figure}
\begin{figure}
\begin{center}
\includegraphics[width=0.9\columnwidth]{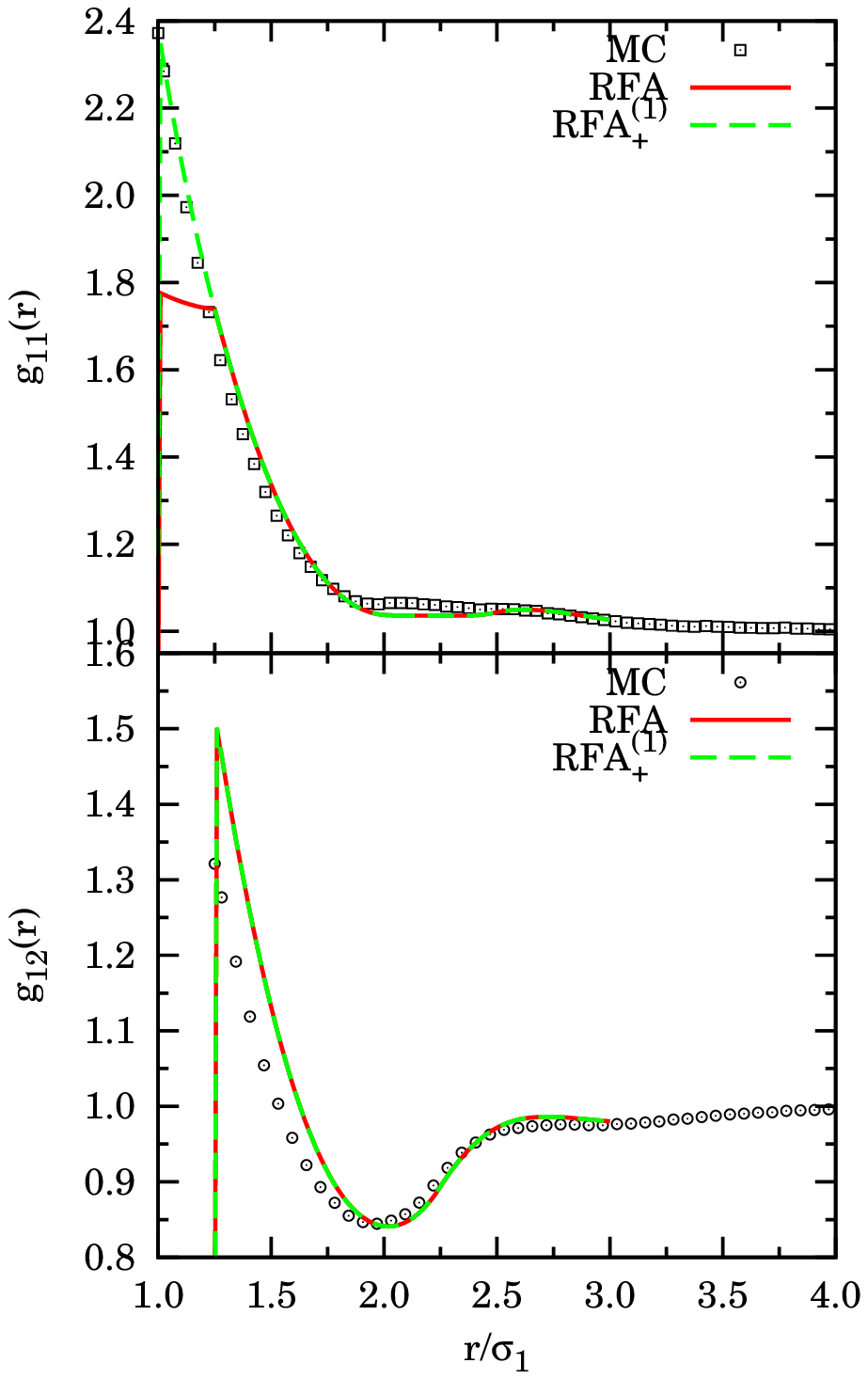}
\end{center}
\caption{(Color online) RDF for system E of table \protect\ref{tab:cases}.}
\label{fig:gr-E1}
\end{figure}
\begin{figure}
\begin{center}
\includegraphics[width=0.9\columnwidth]{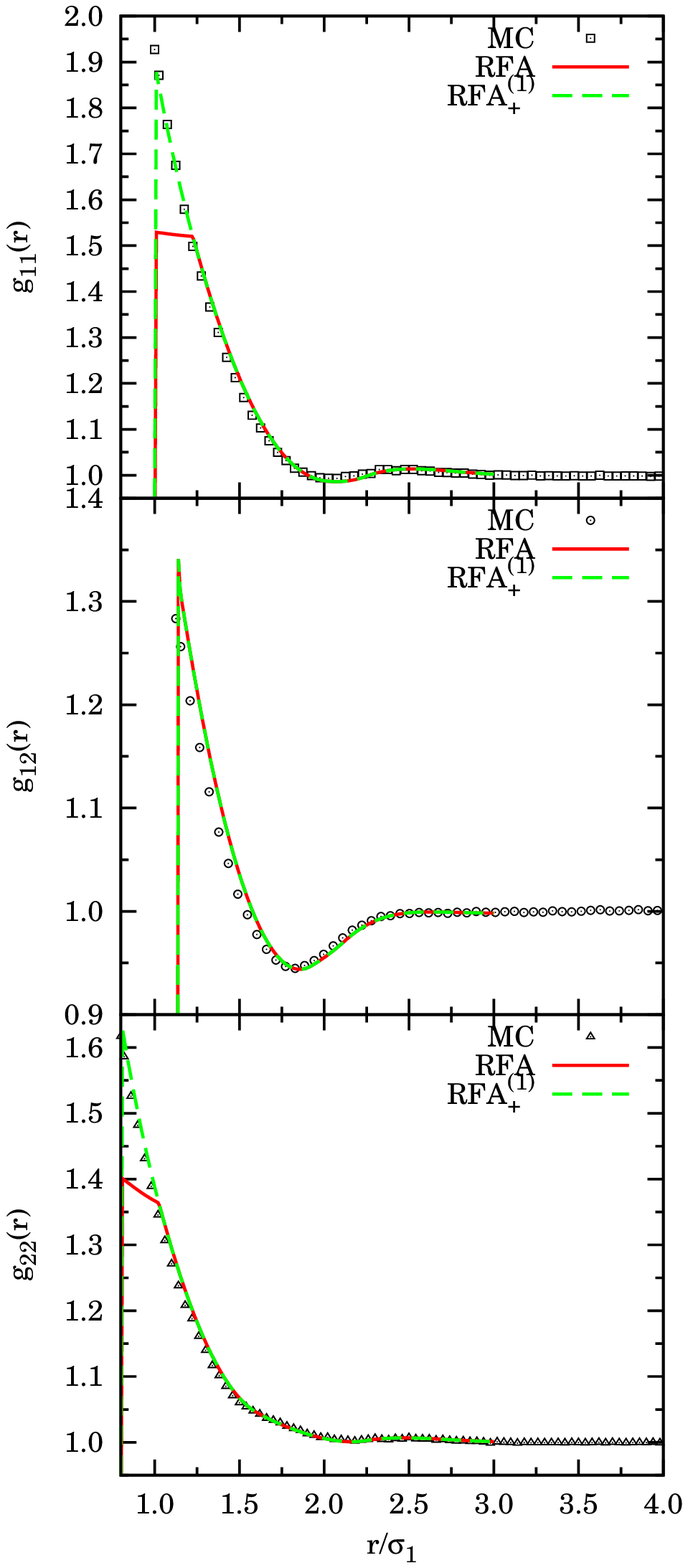}
\end{center}
\caption{(Color online) RDF for system F of table \protect\ref{tab:cases}.}
\label{fig:gr-F1}
\end{figure}
\begin{figure}
\begin{center}
\includegraphics[width=0.9\columnwidth]{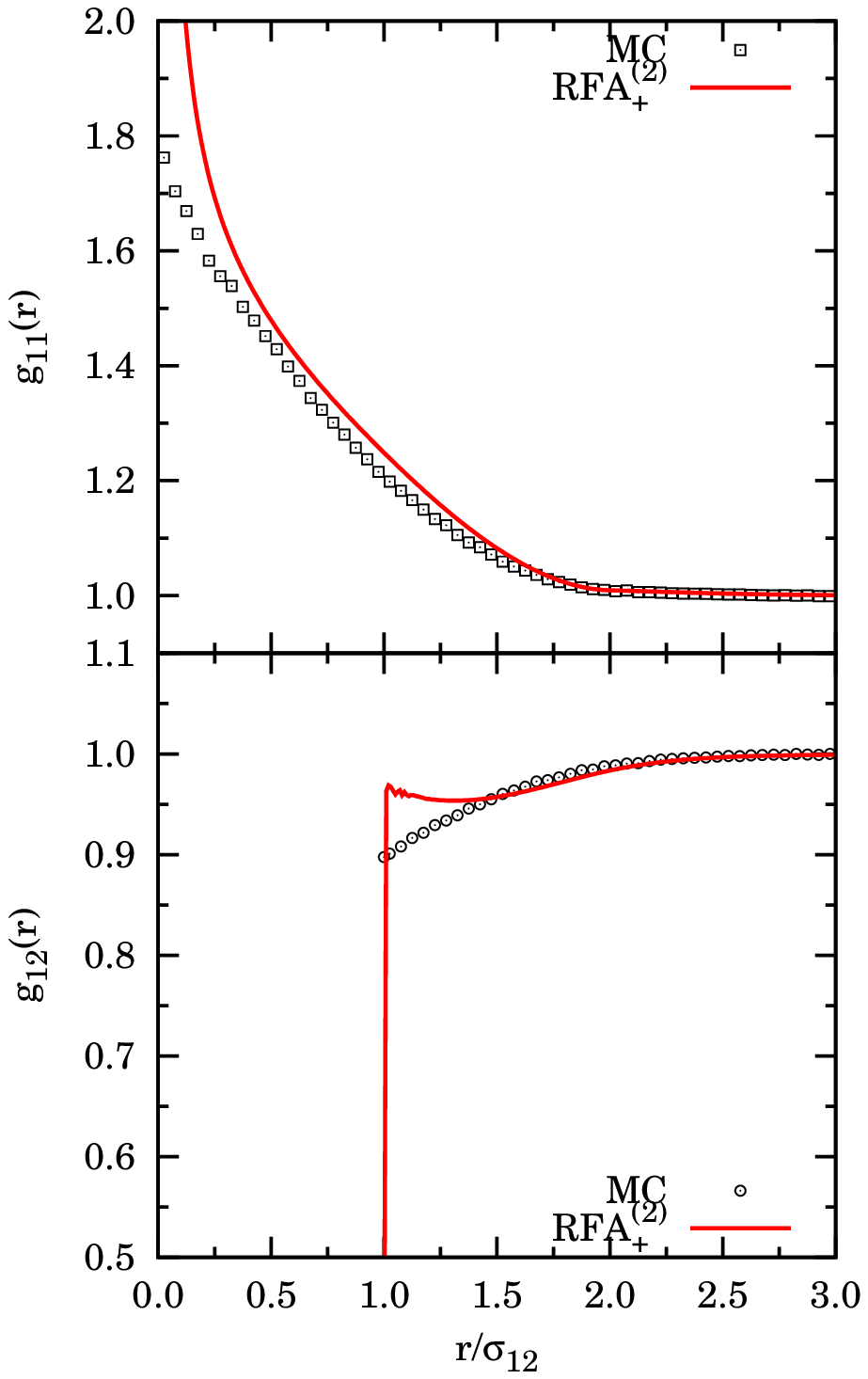}
\end{center}
\caption{(Color online) RDF for the  WR model at $\rho\sigma_{12}^3=0.28748$. The
MC data are taken from Ref.\ \cite{FP04}.
\label{fig:WR1}}
\end{figure}
\begin{figure}
\begin{center}
\includegraphics[width=0.9\columnwidth]{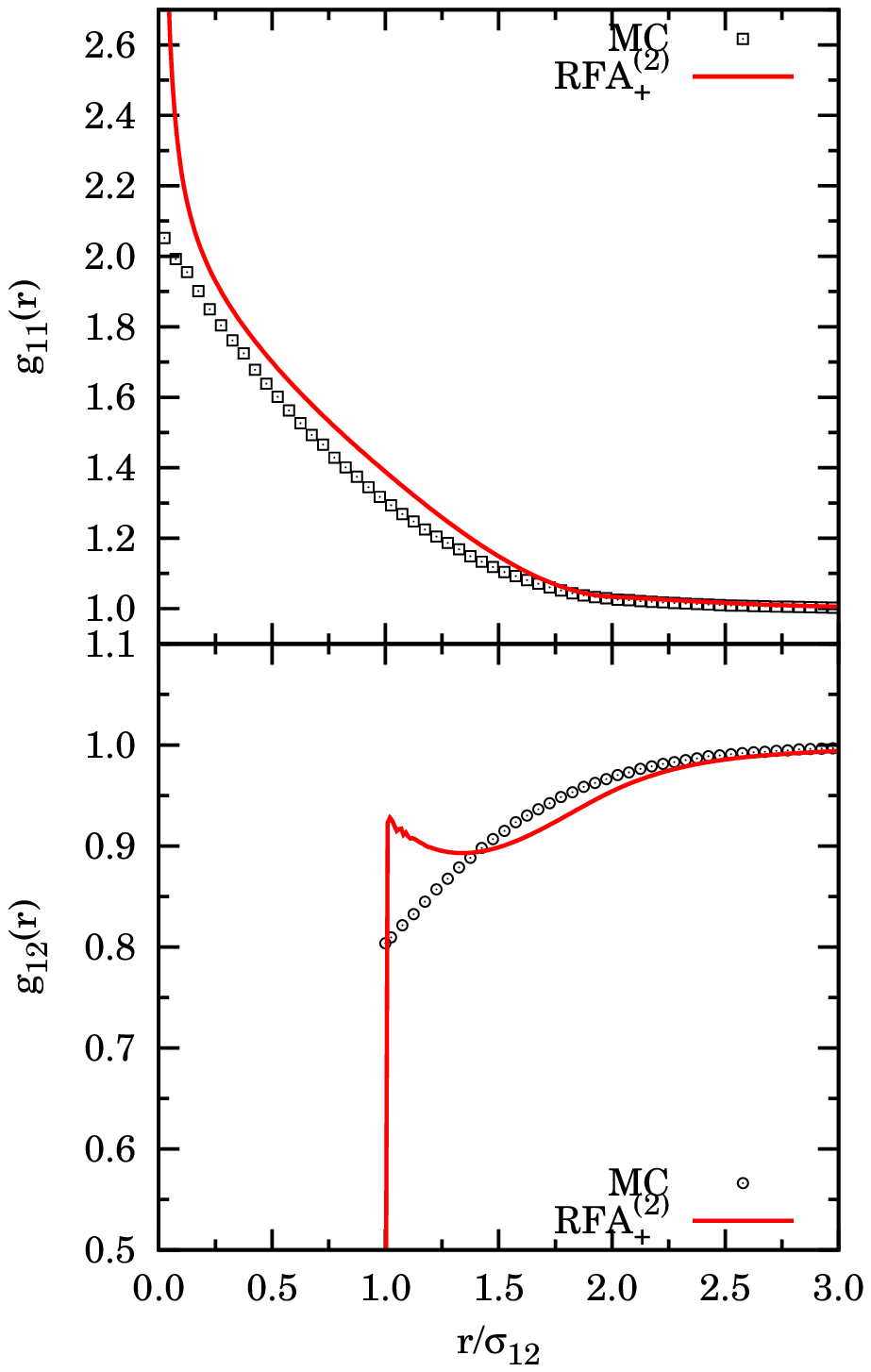}
\end{center}
\caption{(Color online) RDF for the  WR model at $\rho\sigma_{12}^3=0.4$. The
MC data are taken from Ref.\ \cite{FP04}.}
\label{fig:WR2}
\end{figure}

The RDF of approximation $\aA$ is analytically and explicitly given in Laplace space by  Eqs.\ \eqref{15}--\eqref{13NAHS} and \eqref{24}--\eqref{25}.
In real space,  $rg_{ij}(r)$ is easily found by taking the inverse Laplace transform of
$G_{ij}(s)$ through the numerical scheme described in
Ref.\ \cite{AW92}. To get $g_{ij}(r)$ in approximation $\aB^{(m)}$, one needs to make use of Eq.\  \eqref{54},
where $\gamma_{ikj}^{(m)}(r)$ is explicitly given by Eqs.\ \eqref{gamma1} and \eqref{gamma2} for $m=1$ and $m=2$, respectively
\cite{note_11_09}.
Notice that, while the true RDF has to be
symmetric under exchange of species indices, the RDF obtained from
approximation {$\aA$} or $\aB$ is, except for symmetric and equimolar mixtures, not symmetric, i.e., $g_{ij}(r)\neq g_{ji}(r)$ if $i\neq j$. Although this artificial asymmetry is generally small from a practical point of view, it represents a  penalty we pay for our extension of the AHS  solution of the PY equation. To cope with this shortcoming, we  just redefine the like-unlike RDF as the symmetrized one
$\frac{1}{2}[g_{ij}(r)+g_{ji}(r)]$.

In a binary mixture, $\s_{11}=\sigma_{12}+a_{12}=\sigma_1+\frac{1}{2}(\sigma_1+\sigma_2)\Delta$, $\s_{22}=\sigma_{12}-a_{12}=\sigma_2+\frac{1}{2}(\sigma_1+\sigma_2)\Delta$, and $\s_{12}=\frac{1}{2}(\sigma_1+\sigma_2)$. Therefore, $\s_{11}<\sigma_1$ and $\s_{22}<\sigma_2$ for $\Delta<0$, while $\s_{12}<\sigma_{12}$ for $\Delta>0$. In what follows, we will truncate $\left.g_{ij}(r)\right|_{\aA}$ for $r<\sigma_{ij}$ when $\s_{ij}<\sigma_{ij}$.

In order to evaluate the merits and limitations of the structural properties predicted by our approximations, we have performed canonical MC simulations  of the binary NAHS system
with $N=2196$ particles and $10^5N$ MC steps per run. The cell index
method has been used \cite{AT87}.  The statistical error on the RDF is within the size of the symbols
used in the graphs reported.

We have chosen six representative systems, all within the region  $-\sigma_2/(\sigma_1+\sigma_2)\leq\Delta\leq
2\sigma_2/(\sigma_1+\sigma_2)$ assumed in the construction of approximation $\aB$. Those six systems are represented in Fig.\ \ref{diagram2} and their respective values of composition and density are displayed in Table  \ref{tab:cases}. Three of the mixtures have a negative nonadditivity (A, B, and C), while the other three have a positive nonadditivity (D, E, and F). Moreover, there are four equimolar symmetric mixtures (A, B, D, and E) and two asymmetric ones (C and F). In those two latter cases, however, both species contribute almost equally to the (nominal) packing fraction $\eta$ since $x_1\sigma_1^3/x_2\sigma_2^3=(5/4)^3/2=125/128\simeq 0.98$.

\subsection{Negative nonadditivity}
\subsubsection{Symmetric mixtures}

Figures \ref{fig:gr-A} and \ref{fig:gr-B} display the RDF for systems A and B,  respectively. System A is only slightly nonadditive and we observe that both approximations $\aA$ and $\aB^{(1)}$ do a very good job. On the other hand, while $\aA$ and $\aB^{(1)}$ coincide for $g_{11}(r)$ with $r>\sigma_1$, they differ for $g_{12}(r)$ in the interval $\sigma_{12}=0.9\sigma_1\leq r\leq \s_{12}=\sigma_1$. In fact, approximation $\aA$ presents an artificial discontinuity of the first derivative $g_{12}'(r)$ at $r=\sigma_1$. This is corrected by approximation $\aB^{(1)}$, which presents a good agreement with the MC results for $r<\sigma_1$. In spite of this, we observe that approximation $\aB^{(1)}$ underestimates the contact value $g_{12}(\sigma_{12}^+)$, in agreement with the entry of Table \ref{tab:contact} corresponding to case A.

In the case of system B the nonadditivity is larger and, according to Fig.\ \ref{fig:gr-B}, the performance of our approximations is still good for $g_{11}(r)$ but worsens for $g_{12}(r)$. In fact, $\left. g_{12}(r)\right|_{\aA}$  turns out to be better than $\left. g_{12}(r)\right|_{\aB^{(1)}}$ in the region $\sigma_{12}=0.5\sigma_1\leq r\leq \s_{12}=\sigma_1$, in agreement with the entry of Table \ref{tab:contact} corresponding to case B. In any case, it is interesting to remark that approximation $\aB^{(1)}$ succeeds in capturing the non-monotonic behavior of $g_{12}(r)$ very near $r=\sigma_{12}$ observed in the simulations.

\subsubsection{Asymmetric mixture}

The only  case representing an asymmetric mixture with negative nonadditivity (system C) is shown in Fig.\ \ref{fig:gr-C}. Again, the MC like-like RDF are very well reproduced by the two approximations. In the case of the like-unlike function $g_{12}(r)$, approximation $\aB^{(1)}$ clearly improves approximation $\aA$ in the region $\sigma_{12}=0.5\sigma_1\leq r\leq \s_{12}=0.9\sigma_1$. Apart from that, both approximations overestimate $g_{12}(r)$ between $r=\s_{12}=0.9\sigma_1$ and the location of the first minimum at about $r\simeq 1.25\sigma_1$.
In Fig.\ \ref{fig:gr-C} we have taken $g_{12}(r)\to\frac{1}{2}[g_{12}(r)+g_{21}(r)]$, as explained at the beginning of this section. Prior to this symmetrization, the maximum relative deviation between $g_{12}(r)$ and $g_{21}(r)$ occurs at $r\simeq 0.75\sigma_1$ and is less than 5\%.

\subsection{Positive nonadditivity}
\subsubsection{Symmetric mixtures}

Let us consider now positive nonadditivities, starting with symmetric mixtures.
Figures \ref{fig:gr-D} and \ref{fig:gr-E1} show the results for systems D and E, respectively. For a small nonadditivity $\Delta=0.05$, both approximations provide very good results, except for $g_{11}(r)$ near contact (see also Table \ref{tab:contact}). Notice, however, that approximation $\aB^{(1)}$ improves approximation $\aA$ in the narrow region $\sigma_{1}\leq r\leq \s_{11}=1.05\sigma_1$.

For a larger nonadditivity (system E), Fig.\ \ref{fig:gr-E1} shows the excellent job made by approximation $\aB^{(1)}$ in the interval  $\sigma_{1}\leq r\leq \s_{11}=1.25\sigma_1$. In the case of the like-unlike correlation function, however, the approximations overestimate the values between $\sigma_{12}$ and the first minimum ($r\simeq 2\sigma_1$).

\subsubsection{Asymmetric mixture}

Figure \ref{fig:gr-F1} displays the three functions $g_{ij}(r)$ for the asymmetric system F. As in case E, approximation $\aB^{(1)}$  nicely reproduces the exact results from the simulation for the like-like correlations and
corrects the unphysical kink  of approximation {$\aA$} occurring
at $\s_{11}=1.225\sigma_1$ and $\s_{22}=1.025\sigma_1$. Interestingly enough, although the values of $\Delta$ and $\rho\sigma_1^3$ are the same in systems E and F, the performance of the approximations for $g_{12}(r)$ is much better in case F (asymmetric mixture) than in case E (symmetric mixture). This might be partially due to the fact that the  packing fraction $\eta$ is smaller in system F than in system E.
For the asymmetric system F, we have found that the maximum relative deviation between $g_{12}(r)$ and $g_{21}(r)$ takes place at $r=\sigma_{12}=\frac{9}{8}\sigma_1$ and is less than $0.5$\%.

\subsection{The Widom--Rowlinson model}
\label{sec:wr}

As recalled in Sec.\ \ref{sec:introduction}, the WR model corresponds to an equimolar symmetric binary  NAHS mixture where
$\sigma_1=\sigma_2=0$ and $\sigma_{12}\neq 0$. The model is then
fully characterized by  the reduced density, $\rho\sigma_{12}^3$. The
critical demixing reduced density for this model is around $0.75$
\cite{SY96,JGMC97}.

The nonadditivity parameter of the WR model is $\Delta=\sigma_{12}/\sigma_{12}^\add-1\to\infty$, so  it lies  well outside the
``safe'' region for our approximation $\aB$ (see Fig.\ \ref{diagram2}). To compensate for this, we replace here approximation $\aB^\one$ by approximation $\aB^{(2)}$.

We see from Figs.\  \ref{fig:WR1} and \ref{fig:WR2} that approximation $\aB^{(2)}$ does a much better job than expected at the two densities considered. The main drawbacks of the theory are that the contact value  $g_{11}(0)$ is dramatically overestimated and the behavior of $g_{12}(r)$ for $r\gtrsim \sigma_{12}$ is qualitatively wrong.
In spite of this, it is remarkable that approximation $\aB^{(2)}$ captures well the global properties of the RDF in this extreme system.

\section{Summary and conclusions}
\label{sec:conclusions}

The importance of the NAHS model in liquid state theory cannot be overemphasized. When the reference or effective interaction among the microscopic components (at an atomic or a colloidal level of description) of a statistical system is modeled as of hard-core type, there is no reason to expect that the interaction range $\sigma_{ij}$ corresponding to the pair $(i,j)$ is enslaved to be the arithmetic mean of the interaction ranges $\sigma_i$ and $\sigma_j$ corresponding to the pairs $(i,i)$ and $(j,j)$, respectively. Therefore, in an $n$-component NAHS mixture the number of independent interaction ranges is $n(n+1)/2$, in contrast to the number $n$ in an AHS mixture. It is then not surprising that, while an exact solution of the PY  theory exists for AHS systems \cite{L64}, numerical methods are needed when solving the PY and other integral-equation theories for NAHSs \cite{BPGG86}. Therefore, analytical approaches to the problem can represent attractive and welcome contributions.

In this paper we have constructed a non-perturbative fully analytical approximation for the Laplace transforms $G_{ij}(s)$ of $r g_{ij}(r)$, where $g_{ij}(r)$ is the set of RDF of a general 3D NAHS fluid mixture.
Our approach follows several stages. The starting point is the analytical PY solution for AHSs, Eqs.\ \eqref{Grfa}--\eqref{Bdef}. Exploiting the connection between the exact solutions for 1D NAHS and AHS mixtures [see Eqs.\ \eqref{s1} and \eqref{9}], the AHS PY solution is rewritten in an alternative form, Eqs.\ \eqref{11}--\eqref{15}. Our approximation $\aA$ consists of keeping the form \eqref{15}, except that $\deltak_{ij}$ in Eq.\ \eqref{11} is replaced by $\sigma_{ij}$ [cf.\ Eq.\ \eqref{11NAHS}] and $\sigma_i$ in Eq.\ \eqref{13} is replaced by $b_{ij}\equiv \sigma_{ij}+a_{ij}$ [cf.\ Eq.\ \eqref{13NAHS}]. Moreover, the parameters $L_{ij}^\zero$ and $L_{ij}^\one$ are no longer given by Eq.\ \eqref{LijPY} but are determined by enforcing the condition \eqref{cond_ii} or, equivalently, Eq.\ \eqref{s2G}. This results in Eqs.\ \eqref{24}--\eqref{25}, and so the problem becomes completely closed and analytical in Laplace space.
The equation of state is obtained either via the virial route \eqref{Zv} through the contact values \eqref{85} or via the compressibility route \eqref{chih} through the coefficients $H_{ij}^\one$ in the expansion of $s^2 G_{ij}(s)$ in powers of $s$, Eq.\ \eqref{s2G}.

The penalty we pay for ``stretching'' the AHS PY solution to the NAHS domain in the way described above is that  $g_{ij}(r)$ may not be strictly zero for $r<\sigma_{ij}$ or may exhibit first-order discontinuities at artificial distances. To deal with this problem, we have restricted ourselves to mixtures such that the first two singularities of $g_{ij}(r)$ are $\sigma_{ij}$ and $\s_{ij}\equiv\min(\sigma_{ik}-a_{kj};k=1,\ldots, n; k\neq j)$. In the binary case ($n=2$) this restriction corresponds to $-\sigma_2/(\sigma_1+\sigma_2)\leq \Delta \leq 2\sigma_2/(\sigma_1+\sigma_2)$ (see Fig.\ \ref{diagram2}). Next, we have constructed a modified approximation $\aB$ whereby either $g_{ij}(r)$ is  truncated for $r<\sigma_{ij}$ if $\s_{ij}<\sigma_{j}$ or the behavior of $g_{ij}(r)$ for $r\gtrsim \s_{ij}$ is extrapolated to the interval $\sigma_{ij}<r<\s_{ij}$ if $\s_{ij}>\sigma_{j}$ [cf.\ Eq.\ \eqref{52bis}]. {}From a practical point of view, the latter extrapolation can be replaced by a polynomial approximation (e.g., linear or quadratic), yielding approximation $\aB^{(m)}$ [cf.\ Eq.\ \eqref{54}]. This is sufficient to guarantee that the slope of $g_{ij}(r)$ is continuous everywhere for $r>\sigma_{ij}$.

For comparison with MC data of the equation of state we have used approximation $\aA$ since its local limitations at the level of the RDF are largely smoothed out when focusing on the thermodynamic properties. The results show that, if the density is low enough as to make both thermodynamic routes practically coincide, our approximation accurately predicts the MC data, as shown in Fig.\ \ref{fig:Zs} and in the bottom panel of Fig.\ \ref{fig:Zar}. For larger densities, the virial and compressibility routes tend to underestimate and overestimate, respectively, the simulation values, this being a typical PY feature. As in the AHS case, the simple interpolation rule $Z=\frac{1}{3}Z^v+\frac{2}{3}Z^c$ provides very good results, except for large nonadditivities (see Fig.\ \ref{fig:Za} and the top and middle panels of Fig.\ \ref{fig:Zar}).

Regarding the structural properties, approximation $\aB^\one$ is found to perform quite well. The contact values are generally more accurate than those obtained from the numerical solution of the PY integral equation, at least for symmetric mixtures, as shown in Table \ref{tab:contact}. Comparison with our own MC simulations shows a very good agreement, except in the case of the like-unlike RDF for distances smaller than the location of the first minimum for large nonadditivities (cf.\ Figs.\ \ref{fig:gr-A}--\ref{fig:gr-F1}). On the other hand, even in the case of the WR model ($\Delta\to\infty$, well beyond the ``safe'' region of Fig.\ \ref{diagram2}) our approximation $\aB^{(2)}$ does a much better job than expected, as illustrated in Figs.\ \ref{fig:WR1} and \ref{fig:WR2}.

In conclusion, one can reasonably argue that our approximation $\aA$, along with its variants $\aB$ and $\aB^{(m)}$, represent excellent compromises between simplicity and accuracy. We have tried other alternative analytical approaches (simpler as well as  more complex) also based on the PY solution for AHSs but none of them has been found to present a behavior as sound and consistent as those proposed in this paper. We expect that they can be useful in the investigation of such an important statistical-mechanical system (both by itself and also as a reference to other systems) as the NAHS mixture.

{The work presented in this paper can be continued along several lines. In particular, we plan to explore in the near future the predictions for the demixing transition
from our approximations. It is also worth exploring the NAHS theory that arises when the starting point is not the PY solution for AHSs but the more advanced RFA proposed in Ref.\ \cite{YSH98}, which contains free parameters that can be accommodated to fit any desired EOS in a thermodynamically consistent way.}

\begin{acknowledgments}
The MC simulations presented in Sec.\ \ref{sec:structure}  were carried out at the Center for High Performance Computing (CHPC), CSIR Campus, 15 Lower Hope St., Rosebank, Cape Town, South Africa.
RF  acknowledges the kind hospitality of the Department
of Physics of the University of Extremadura at Badajoz. The research  of AS has been supported by the Ministerio de  Ciencia e Innovaci\'on (Spain) through Grant No.\ {FIS2010-16587} and  the Junta de Extremadura (Spain) through Grant No.\ GR10158, partially financed by FEDER funds.
\end{acknowledgments}

\appendix

\section{Explicit expressions of $G_{ij}(s)$ for binary mixtures in
   approximation $\aA$}
\label{app0}
By performing the inversion of the matrix \eqref{Qij} and carrying out
the matrix product in Eq.\ \eqref{15} one gets
\begin{widetext}
  \beqa
G_{11}(s)&=&\frac{s^{-2}}{D(s)}\left\{L_{11}(s)\left[1-\frac{2\pi\rho
    x_2}{s^3}N_{22}(s)\right]e^{-\sigma_1 s}\right.
+\frac{2\pi\rho
  x_2}{s^3}L_{11}(s)L_{22}(s)e^{-(\sigma_1+\sigma_2) s}
\nn
&&-\frac{2\pi\rho x_2}{s^3}L_{12}(s)L_{21}(s)e^{-2\sigma_{12} s}
\left.
+\frac{2\pi\rho x_2}{s^3}L_{12}(s)N_{21}(s)e^{-(\sigma_{12}+a_{12}) s}\right\},
\label{60}
\eeqa
\beq
G_{12}(s)=\frac{s^{-2}}{D(s)}\left\{L_{12}(s)\left[1-\frac{2\pi\rho
    x_1}{s^3}N_{11}(s)\right]e^{-\sigma_{12} s}
+\frac{2\pi\rho x_1}{s^3}L_{11}(s)N_{12}(s)e^{-(\sigma_{1}+\sigma_{2})
  s/2}\right\},
\label{61}
\eeq
where the quadratic functions $N_{kj}(s)$ can be found in Eq.\ \eqref{Nkj}
and
\beqa
D(s)&=&\left[1-\frac{2\pi\rho
    x_1}{s^3}N_{11}(s)\right]\left[1-\frac{2\pi\rho
    x_2}{s^3}N_{22}(s)\right]-\frac{(2\pi\rho)^2
  x_1x_2}{s^6}N_{12}(s)N_{21}(s)
\nn
&&+\frac{2\pi\rho x_1}{s^3}L_{11}(s)\left[1-\frac{2\pi\rho x_2}{s^3}
  N_{22}(s)\right]e^{-\sigma_1 s}
+\frac{2\pi\rho x_2}{s^3}L_{22}(s)\left[1-\frac{2\pi\rho x_1}{s^3}
  N_{11}(s)\right]e^{-\sigma_2 s}\nn
&&+\frac{4\pi^2\rho x_1x_2}{s^6}\left[L_{11}(s)L_{22}(s)
  e^{-(\sigma_1+\sigma_2) s}
-L_{12}(s)L_{21}(s)e^{-2\sigma_{12} s}\right.\nn
&&\left.+L_{12}(s)N_{21}(s)e^{-(\sigma_{12}+a_{12}) s}+
L_{21}(s)N_{12}(s)e^{-(\sigma_{12}-a_{12}) s}\right]
\label{62}
\eeqa
\end{widetext}
is the determinant of the matrix $\mathsf{Q}^{-1}$. The expressions
for $G_{22}(s)$ and $G_{21}(s)$ can be obtained by the exchange
$1\leftrightarrow 2$.

\section{Ordering of singular distances in approximation {$\aA$} for binary mixtures}
\label{app1}
{By ``singular'' distances we will refer to} those values of $r$ where the RDF $g_{ij}(r)$ or any of its derivatives have a discontinuity. Physical singularities are located, for instance,  at $r=\sigma_{ij}$ and $r=\sigma_{ik}+\sigma_{kj}$, $k=1,\ldots, n$. Apart from that, approximation {$\aA$} introduces spurious singularities at other distances.

Let us particularize to a binary mixture. The physical \emph{leading} singularity of $g_{ij}(r)$ should be located at $r=\sigma_{ij}$. However, according  to  Eq.\ \eqref{60}, the {leading} singularity of $g_{11}(r)$ takes place at $r=\text{min}(\sigma_1,\sigma_{12}+a_{12},2\sigma_{12})$.  Analogously, the leading singularity of $g_{22}(r)$ is located at $r=\text{min}(\sigma_2,\sigma_{12}-a_{12},2\sigma_{12})$. Finally, Eq.\ \eqref{61} shows that the leading singularity of $g_{12}(r)$ is $r=\frac{1}{2}\text{min}(2\sigma_{12},\sigma_{1}+\sigma_{2})$.
{Note that we have assumed $\sigma_{12}-a_{12}>0$, so that the denominator $D(s)$, Eq.\ \eqref{62}, does not affect the leading singularity of $g_{ij}(r)$}.

\begin{figure}
\includegraphics[width=.9\columnwidth]{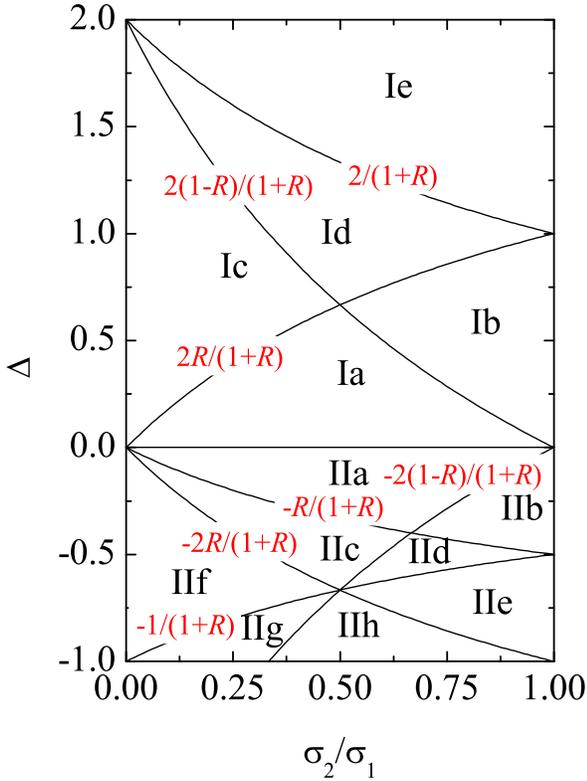}
\caption{(Color online) Plane $\Delta$ vs $R\equiv \sigma_2/\sigma_1$ showing the
  regions with different ordering of the distances $\sigma_1$, $\sigma_2$,     $\sigma_{12}-a_{12}$,  $\sigma_{12}+a_{12}$, $2\sigma_{12}$, and $\sigma_1+\sigma_2$. \label{diagram1}}
\end{figure}

It is thus important to determine the relative ordering of the values $\sigma_1$, $\sigma_2$,     $\sigma_{12}-a_{12}$,  $\sigma_{12}+a_{12}$, $2\sigma_{12}$, and $\sigma_1+\sigma_2$.
Such an ordering depends on the values of $\Delta$ and
$R\equiv\sigma_2/\sigma_1$, where, without loss of generality, we assume that
$\sigma_2\leq\sigma_1$.  A detailed analysis shows that the
$\Delta$-$R$ plane  can be split into 13 disjoint regions with
distinct order for the above singular distances. Those regions are
indicated  in Fig.\ \ref{diagram1}, while Table \ref{tab1} shows the
order applying within each region.
Note that $\sigma_{12}-a_{12}$ is negative in Regions IIf, IIg, and
IIh, i.e., if $-1\leq\Delta\leq -2R/(1+R)$, {thus invalidating those regions from the preceding analysis}.

\begin{table*}
\caption{Order of the singular distances $\sigma_1$, $\sigma_2$,     $\sigma_{12}-a_{12}$,  $\sigma_{12}+a_{12}$, $2\sigma_{12}$, and $\sigma_1+\sigma_2$ in each of the regions
  of Fig.\ \protect\ref{diagram1}.
\label{tab1}}
\begin{ruledtabular}
\begin{tabular} {lccccccccccccc}
Region& \multicolumn{13}{c}{Order}\\
\hline
Ia&$0$&$\leq$&$\sigma_2$&$\leq$&$\sigma_{12}-a_{12}$&$\leq$&$\sigma_1$&$\leq$&$\sigma_{12}+a_{12}$&$\leq$&$\sigma_1+\sigma_2$&$\leq
$&$2\sigma_{12}$\\
Ib&$ 0$&$\leq$&$\sigma_2$&$\leq$&$\sigma_1$&$\leq$&$\sigma_{12}-a_{12}$&$\leq$&$\sigma_{12}+a_{12}$&$\leq$&$\sigma_1+\sigma_2$&$\leq $&$2\sigma_{12}$\\
Ic&$0$&$\leq$&$\sigma_2$&$\leq$&$\sigma_{12}-a_{12}$&$\leq$&$\sigma_1$&$\leq$&$\sigma_1+\sigma_2$&$\leq$&$\sigma_{12}+a_{12}$&$\leq$&$ 2\sigma_{12}$\\
Id&$0$&$\leq$&$\sigma_2$&$\leq$&$\sigma_1$&$\leq$&$\sigma_{12}-a_{12}$&$\leq$&$\sigma_1+\sigma_2$&$\leq$&$\sigma_{12}+a_{12}$&$\leq$&$ 2\sigma_{12}$\\
Ie&$ 0$&$\leq$&$\sigma_2$&$\leq$&$\sigma_1$&$\leq$&$\sigma_1+\sigma_2$&$\leq$&$\sigma_{12}-a_{12}$&$\leq$&$\sigma_{12}+a_{12}$&$\leq$&$ 2\sigma_{12}$\\
IIa&$0$&$\leq$&$\sigma_{12}-a_{12}$&$\leq$&$\sigma_{2}$&$\leq$&$\sigma_{12}+a_{12}$&$\leq$&$\sigma_1$&$\leq$&$2\sigma_{12}$&$\leq
$&$\sigma_1+\sigma_2$\\
IIb&$0$&$\leq$&$\sigma_{12}-a_{12}$&$\leq$&$\sigma_{12}+a_{12}$&$\leq$&$\sigma_{2}$&$\leq$&$\sigma_1$&$\leq$&$2\sigma_{12}$&$\leq
$&$\sigma_1+\sigma_2$\\
IIc&$0$&$\leq$&$\sigma_{12}-a_{12}$&$\leq$&$\sigma_{2}$&$\leq$&$\sigma_{12}+a_{12}$&$\leq$&$2\sigma_{12}$&$\leq$&$\sigma_1$&$\leq
$&$\sigma_1+\sigma_2$\\
IId&$0$&$\leq$&$\sigma_{12}-a_{12}$&$\leq$&$\sigma_{12}+a_{12}$&$\leq$&$\sigma_{2}$&$\leq$&$2\sigma_{12}$&$\leq$&$\sigma_1$&$\leq
$&$\sigma_1+\sigma_2$\\
IIe&$0$&$\leq$&$\sigma_{12}-a_{12}$&$\leq$&$\sigma_{12}+a_{12}$&$\leq$&$2\sigma_{12}$&$\leq$&$\sigma_{2}$&$\leq$&$\sigma_1$&$\leq
$&$\sigma_1+\sigma_2$\\
IIf&$\sigma_{12}-a_{12}$&$\leq$&$0$&$\leq$&$\sigma_{2}$&$\leq$&$2\sigma_{12}$&$\leq$&$\sigma_{12}+a_{12}$&$\leq$&$\sigma_1$&$\leq
$&$\sigma_1+\sigma_2$\\
IIg&$\sigma_{12}-a_{12}$&$\leq$&$0$&$\leq$&$2\sigma_{12}$&$\leq$&$\sigma_{2}$&$\leq$&$\sigma_{12}+a_{12}$&$\leq$&$\sigma_1$&$\leq
$&$\sigma_1+\sigma_2$\\
IIh&$\sigma_{12}-a_{12}$&$\leq$&$0$&$\leq$&$2\sigma_{12}$&$\leq$&$\sigma_{12}+a_{12}$&$\leq$&$\sigma_{2}$&$\leq$&$\sigma_1$&$\leq
$&$\sigma_1+\sigma_2$\\
\end{tabular}
\end{ruledtabular}
\end{table*}

We observe that $\sigma_1$ and $\sigma_2$ are indeed the leading singularities of $g_{11}(r)$ and $g_{22}(r)$, respectively, for positive nonadditivity (regions Ia--Ie). Reciprocally, $\sigma_{12}$ is the leading singularity of $g_{12}(r)$ for negative nonadditivity (regions IIa--IIh).

In order to construct approximation $\aB$, we want to restrict ourselves to those regions such that the two leading singularities of $g_{11}(r)$ are $\sigma_1$ and $\s_{11}\equiv \sigma_{12}+a_{12}$. Inspection of Table \ref{tab1} shows that Regions IIc--IIh are discarded by this criterion. In the remaining regions the leading singularity of $g_{11}(r)$ is $\text{min}(\sigma_1,\sigma_{12}+a_{12})$ but the next one is not necessarily $\text{max}(\sigma_1,\sigma_{12}+a_{12})$ since the latter value competes with $\text{min}(\sigma_1,\sigma_{12}+a_{12})+\text{min}(\sigma_2,\sigma_{12}-a_{12},2\sigma_{12})$, where the term $\text{min}(\sigma_2,\sigma_{12}-a_{12},2\sigma_{12})$ comes from the denominator $D(s)$ [cf.\ Eq.\ \eqref{62}]. It can be checked that $\text{max}(\sigma_1,\sigma_{12}+a_{12})\geq \text{min}(\sigma_1,\sigma_{12}+a_{12})+\text{min}(\sigma_2,\sigma_{12}-a_{12},2\sigma_{12})$ in Regions Ic--Ie. Therefore the two first singularities of $g_{11}(r)$ are $\sigma_1$ and $\s_{11}=\sigma_{12}+a_{12}$ in Regions Ia, Ib, IIa, and IIb only. It turns out that in those four regions  the two leading singularities of $g_{22}(r)$ are $\sigma_2$ and $\s_{22}\equiv \sigma_{12}-a_{12}$, and the two leading singularities of $g_{12}(r)$ are $\sigma_{12}$ and $\s_{12}\equiv \frac{1}{2}(\sigma_{1}+\sigma_{2})$.

In summary, Regions Ia, Ib, IIa, and IIb are the only ones where the two leading singularities of $g_{ij}(r)$ are $\sigma_{ij}$ and $\s_{ij}\equiv\sigma_{ik}-a_{kj}$ with $k\neq j$.

\section{Short-range forms of  $g_{ij}(r)$ for binary mixtures in
  approximation $\aA$}
\label{app2}
In what follows we assume that $-\sigma_2/(\sigma_1+\sigma_2)\leq \Delta\leq 2\sigma_2/(\sigma_1+\sigma_2)$, which corresponds to Regions Ia, Ib, IIa, and IIb of Fig.\ \ref{diagram1}. As discussed in Appendix \ref{app1}, this guarantees that the first two singularities of $g_{ij}(r)$ are $\sigma_{ij}$ and $\s_{ij}\equiv \sigma_{ik}-a_{kj}$ with $k\neq j$. The aim of this Appendix is to give the expressions of $g_{ij}(r)$ in the region $0\leq r\leq \max(\sigma_{ij},\s_{ij})+\epsilon$, where $\epsilon$ is  smaller than the separation between $\max(\sigma_{ij},\s_{ij})$ and the next singularity.

It is convenient to  assign a bookkeeping parameter $z$ to $e^{-s}$,
so that, for instance, $e^{-\sigma_{ij}s}$ becomes $z^{\sigma_{ij}}e^{-\sigma_{ij}s}$.
We will set $z=1$ at the end of the calculations. Therefore, the denominator $D(s)$ given by Eq.\ \eqref{62} becomes
\beq
D(s)=D_0(s)+o(z^0),
\label{63a}
\eeq
where
\beqa
D_0(s)&=&\left[1-\frac{2\pi\rho
    x_1}{s^3}N_{11}(s)\right]\left[1-\frac{2\pi\rho
    x_2}{s^3}N_{22}(s)\right]\nn
    &&-\frac{(2\pi\rho)^2
  x_1x_2}{s^6}N_{12}(s)N_{21}(s).
\label{57}
\eeqa
In Eq.\ \eqref{63a},
$o(z^n)$ denotes terms that are negligible versus $z^n$ in the
(formal) limit $z\to 0$, i.e., $\lim_{z\to 0} z^{-n} o(z^n)=0$.
{}From Eq.\ \eqref{60} we see that the two leading terms in $G_{11}(s)$
are of orders $z^{\sigma_1}$ and $z^{\sigma_{12}+a_{12}}$:
\beqa
G_{11}(s)&=&\Phi_{11}(s)e^{-\sigma_1 s}z^{\sigma_1}
+{2\pi\rho x_2}\Gamma_{121}(s)e^{-(\sigma_{12}+a_{12})
  s}\nn
  &&\times z^{\sigma_{12}+a_{12}}+o(z^{\sigma_1})+o(z^{\sigma_{12}+a_{12}}),
\label{63}
\eeqa
where
\beq
\Phi_{11}(s)\equiv
\frac{s^{-2}}{D_0(s)}L_{11}(s)\left[1-\frac{2\pi\rho
    x_2}{s^3}N_{22}(s)\right]
\label{64}
\eeq
and $\Gamma_{ikj}(s)$ is given by Eq.\ \eqref{55}.
Analogously,
\beqa
G_{12}(s)&=&\Phi_{12}(s)e^{-\sigma_{12} s}z^{\sigma_{12}}
+{2\pi\rho x_1}\Gamma_{112}(s)e^{-(\sigma_{1}+\sigma_{2})
  s/2}\nn
&&\times z^{(\sigma_{1}+\sigma_{2})/2}+o(z^{\sigma_{12}})+
o(z^{(\sigma_{1}+\sigma_{2})/2}),
\label{65}
\eeqa
\beqa
G_{21}(s)&=&\Phi_{21}(s)e^{-\sigma_{12} s}z^{\sigma_{12}}
+{2\pi\rho x_2}\Gamma_{221}(s)e^{-(\sigma_{1}+\sigma_{2}) s/2}
\nn
&&\times z^{(\sigma_{1}+\sigma_{2})/2}+o(z^{\sigma_{12}})+
o(z^{(\sigma_{1}+\sigma_{2})/2}),
\label{66}
\eeqa
\beqa
G_{22}(s)&=&\Phi_{22}(s)e^{-\sigma_2 s}z^{\sigma_2}
+{2\pi\rho x_1}\Gamma_{212}(s)e^{-(\sigma_{12}-a_{12})
  s}\nn
&&\times z^{\sigma_{12}-a_{12}}+o(z^{\sigma_2})+o(z^{\sigma_{12}-a_{12}}),
\label{67}
\eeqa
where
\beq
\Phi_{12}(s)\equiv \frac{s^{-2}}{D_0(s)}L_{12}(s)
\left[1-\frac{2\pi\rho x_1}{s^3}N_{11}(s)\right],
\label{68}
\eeq
\beq
\Phi_{21}(s)\equiv \frac{s^{-2}}{D_0(s)}L_{21}(s)
\left[1-\frac{2\pi\rho x_2}{s^3}N_{22}(s)\right],
\label{69}
\eeq
\beq
\Phi_{22}(s)\equiv \frac{s^{-2}}{D_0(s)}L_{22}(s)
\left[1-\frac{2\pi\rho x_1}{s^3}N_{11}(s)\right].
\label{70}
\eeq

Laplace inversion of Eqs.\ \eqref{63} and
\eqref{65}--\eqref{67} shows that in the interval $0\leq r\leq
\max(\sigma_{ij},\s_{ij})+\epsilon$ we obtain
\beqa
g_{11}(r)&=&\frac{1}{r}\Theta(r-\sigma_1)\phi_{11}
(r-\sigma_1)+\frac{2\pi\rho
  x_2}{r}\Theta(r-\sigma_{12}-a_{1 2})
\nn &&\times
\gamma_{121}(r-\sigma_{12}-a_{12}),
\label{71}
\eeqa
\beqa
g_{12}(r)&=&\frac{1}{r}\Theta(r-\sigma_{12})\phi_{12}
(r-\sigma_{12})+\frac{2\pi\rho
  x_1}{r}\Theta(r-\frac{\sigma_{1}+\sigma_{2}}{2})
  \nn &&\times
\gamma_{112}(r-\frac{\sigma_{1}+\sigma_{2}}{2}),
\label{72}
\eeqa
\beqa
g_{21}(r)&=&\frac{1}{r}\Theta(r-\sigma_{21})\phi_{21}
(r-\sigma_{12})+\frac{2\pi\rho
  x_2}{r}\Theta(r-\frac{\sigma_{1}+\sigma_{2}}{2})
\nn &&\times
\gamma_{221}(r-\frac{\sigma_{1}+\sigma_{2}}{2}),
\label{73}
\eeqa
\beqa
g_{22}(r)&=&\frac{1}{r}\Theta(r-\sigma_2)\phi_{22}
(r-\sigma_2)+\frac{2\pi\rho
  x_1}{r}\Theta(r-\sigma_{12}+a_{12})
  \nn &&\times
  \gamma_{212}(r-\sigma_{12}+a_{12}),
\label{74}
\eeqa
where we have already set $z=1$. In Eqs.\ \eqref{71}--\eqref{74}, $\phi_{ij}(r)$ and $\gamma_{ikj}(r)$ are the inverse Laplace transforms of
$\Phi_{ij}(s)$ and $\Gamma_{ikj}(s)$, respectively.

Since
$\phi_{ij}(0)=\lim_{s\to\infty}\Phi_{ij}(s)=L_{ij}^\one$,  the contact values in approximation {$\aA$} are
\beq
g_{11}(\sigma_1^+)=\frac{L_{11}^\one}{\sigma_1},
\label{75}
\eeq
\beq
g_{12}(\sigma_{12}^+)=\frac{L_{12}^\one}{\sigma_{12}}+\frac{2\pi\rho x_1}{\sigma_{12}}
  \gamma_{112}\left(\sigma_{12}-\frac{\sigma_{1}+\sigma_{2}}{2}\right),
\label{76}
\eeq
\beq
g_{21}(\sigma_{12}^+)=\frac{L_{21}^\one}{\sigma_{12}}+\frac{2\pi\rho x_2}{\sigma_{12}}
  \gamma_{221}\left(\sigma_{12}-\frac{\sigma_{1}+\sigma_{2}}{2}\right),
\label{77}
\eeq
\beq
g_{22}(\sigma_2^+)=\frac{L_{22}^\one}{\sigma_2},
\label{78}
\eeq
in Regions Ia and Ib ($\Delta>0$). On the other hand, in Regions IIa and IIb ($\Delta<0$),
\beq
g_{11}(\sigma_{1}^+)=\frac{L_{11}^\one}{\sigma_{1}}+\frac{2\pi\rho x_2}{\sigma_{1}}
  \gamma_{121}\left(\sigma_{1}-\sigma_{12}-a_{12}\right),
\label{79}
\eeq
\beq
g_{12}(\sigma_{12}^+)=\frac{L_{12}^\one}{\sigma_{12}},
\label{80}
\eeq
\beq
g_{21}(\sigma_{12}^+)=\frac{L_{21}^\one}{\sigma_{12}},
\label{81}
\eeq
\beq
g_{22}(\sigma_{2}^+)=\frac{L_{22}^\one}{\sigma_{2}}+\frac{2\pi\rho x_1}{\sigma_{2}}
  \gamma_{212}\left(\sigma_{2}-\sigma_{12}+a_{12}\right).
\label{82}
\eeq

A more compact form is provided by Eq.\ \eqref{85}.

\bibliographystyle{apsrev}
\bibliography{D:/bib_files/liquid}

\begin{thebibliography}{48}
\expandafter\ifx\csname natexlab\endcsname\relax\def\natexlab#1{#1}\fi
\expandafter\ifx\csname bibnamefont\endcsname\relax
  \def\bibnamefont#1{#1}\fi
\expandafter\ifx\csname bibfnamefont\endcsname\relax
  \def\bibfnamefont#1{#1}\fi
\expandafter\ifx\csname citenamefont\endcsname\relax
  \def\citenamefont#1{#1}\fi
\expandafter\ifx\csname url\endcsname\relax
  \def\url#1{\texttt{#1}}\fi
\expandafter\ifx\csname urlprefix\endcsname\relax\def\urlprefix{URL }\fi
\providecommand{\bibinfo}[2]{#2}
\providecommand{\eprint}[2][]{\url{#2}}

\bibitem[{\citenamefont{Hansen and McDonald}(2006)}]{HM06}
\bibinfo{author}{\bibfnamefont{J.-P.} \bibnamefont{Hansen}} \bibnamefont{and}
  \bibinfo{author}{\bibfnamefont{I.~R.} \bibnamefont{McDonald}},
  \emph{\bibinfo{title}{{Theory of Simple Liquids}}}
  (\bibinfo{publisher}{Academic Press}, \bibinfo{address}{London},
  \bibinfo{year}{2006}).

\bibitem[{\citenamefont{Likos}(2001)}]{L01}
\bibinfo{author}{\bibfnamefont{C.~N.} \bibnamefont{Likos}},
  \bibinfo{journal}{Phys. Rep.} \textbf{\bibinfo{volume}{348}},
  \bibinfo{pages}{267} (\bibinfo{year}{2001}).

\bibitem[{\citenamefont{Mulero}(2008)}]{M08}
\bibinfo{editor}{\bibfnamefont{A.}~\bibnamefont{Mulero}}, ed.,
  \emph{\bibinfo{title}{Theory and Simulation of Hard-Sphere Fluids and Related
  Systems}} (\bibinfo{publisher}{Springer}, \bibinfo{address}{Berlin},
  \bibinfo{year}{2008}), vol. \bibinfo{volume}{753} of
  \emph{\bibinfo{series}{Lectures Notes in Physics}}.

\bibitem[{\citenamefont{Ballone et~al.}(1986)\citenamefont{Ballone, Pastore,
  Galli, and Gazzillo}}]{BPGG86}
\bibinfo{author}{\bibfnamefont{P.}~\bibnamefont{Ballone}},
  \bibinfo{author}{\bibfnamefont{G.}~\bibnamefont{Pastore}},
  \bibinfo{author}{\bibfnamefont{G.}~\bibnamefont{Galli}}, \bibnamefont{and}
  \bibinfo{author}{\bibfnamefont{D.}~\bibnamefont{Gazzillo}},
  \bibinfo{journal}{Mol. Phys.} \textbf{\bibinfo{volume}{59}},
  \bibinfo{pages}{275} (\bibinfo{year}{1986}).

\bibitem[{\citenamefont{Gazzillo et~al.}(1989)\citenamefont{Gazzillo, Pastore,
  and Enzo}}]{GPE89}
\bibinfo{author}{\bibfnamefont{D.}~\bibnamefont{Gazzillo}},
  \bibinfo{author}{\bibfnamefont{G.}~\bibnamefont{Pastore}}, \bibnamefont{and}
  \bibinfo{author}{\bibfnamefont{S.}~\bibnamefont{Enzo}}, \bibinfo{journal}{J.
  Phys.: Condens Matter} \textbf{\bibinfo{volume}{1}}, \bibinfo{pages}{3469}
  (\bibinfo{year}{1989}).

\bibitem[{\citenamefont{Gazzillo et~al.}(1990)\citenamefont{Gazzillo, Pastore,
  and Frattini}}]{GPF90}
\bibinfo{author}{\bibfnamefont{D.}~\bibnamefont{Gazzillo}},
  \bibinfo{author}{\bibfnamefont{G.}~\bibnamefont{Pastore}}, \bibnamefont{and}
  \bibinfo{author}{\bibfnamefont{R.}~\bibnamefont{Frattini}},
  \bibinfo{journal}{J. Phys.: Condens Matter} \textbf{\bibinfo{volume}{2}},
  \bibinfo{pages}{3469} (\bibinfo{year}{1990}).

\bibitem[{\citenamefont{Shouten}(1989)}]{S89}
\bibinfo{author}{\bibfnamefont{J.~A.} \bibnamefont{Shouten}},
  \bibinfo{journal}{Phys. Rep.} \textbf{\bibinfo{volume}{172}},
  \bibinfo{pages}{33} (\bibinfo{year}{1989}).

\bibitem[{\citenamefont{Gast et~al.}(1983)\citenamefont{Gast, Hall, and
  Russel}}]{GHR83}
\bibinfo{author}{\bibfnamefont{A.~P.} \bibnamefont{Gast}},
  \bibinfo{author}{\bibfnamefont{C.~K.} \bibnamefont{Hall}}, \bibnamefont{and}
  \bibinfo{author}{\bibfnamefont{W.~B.} \bibnamefont{Russel}},
  \bibinfo{journal}{J. Colloid Interface Sci.} \textbf{\bibinfo{volume}{96}},
  \bibinfo{pages}{251} (\bibinfo{year}{1983}).

\bibitem[{\citenamefont{Lekkerkerker et~al.}(1992)\citenamefont{Lekkerkerker,
  Poon, Pusey, Stroobants, and Warren}}]{LPPSW92}
\bibinfo{author}{\bibfnamefont{H.~N.~W.} \bibnamefont{Lekkerkerker}},
  \bibinfo{author}{\bibfnamefont{W.~K.} \bibnamefont{Poon}},
  \bibinfo{author}{\bibfnamefont{P.~N.} \bibnamefont{Pusey}},
  \bibinfo{author}{\bibfnamefont{A.}~\bibnamefont{Stroobants}},
  \bibnamefont{and} \bibinfo{author}{\bibfnamefont{P.~B.}
  \bibnamefont{Warren}}, \bibinfo{journal}{Europhys. Lett.}
  \textbf{\bibinfo{volume}{20}}, \bibinfo{pages}{559} (\bibinfo{year}{1992}).

\bibitem[{\citenamefont{Dijkstra et~al.}(1999)\citenamefont{Dijkstra, Brader,
  and Evans}}]{DBE99}
\bibinfo{author}{\bibfnamefont{M.}~\bibnamefont{Dijkstra}},
  \bibinfo{author}{\bibfnamefont{J.~M.} \bibnamefont{Brader}},
  \bibnamefont{and} \bibinfo{author}{\bibfnamefont{R.}~\bibnamefont{Evans}},
  \bibinfo{journal}{J. Phys.: Condens. Matter} \textbf{\bibinfo{volume}{11}},
  \bibinfo{pages}{10079} (\bibinfo{year}{1999}).

\bibitem[{\citenamefont{Meijer and Frenkel}(1994)}]{MF94}
\bibinfo{author}{\bibfnamefont{E.~J.} \bibnamefont{Meijer}} \bibnamefont{and}
  \bibinfo{author}{\bibfnamefont{D.}~\bibnamefont{Frenkel}},
  \bibinfo{journal}{J. Chem. Phys.} \textbf{\bibinfo{volume}{100}},
  \bibinfo{pages}{6873} (\bibinfo{year}{1994}).

\bibitem[{\citenamefont{Santos et~al.}(2005)\citenamefont{Santos, {L\'opez de
  Haro}, and Yuste}}]{SHY05}
\bibinfo{author}{\bibfnamefont{A.}~\bibnamefont{Santos}},
  \bibinfo{author}{\bibfnamefont{M.}~\bibnamefont{{L\'opez de Haro}}},
  \bibnamefont{and} \bibinfo{author}{\bibfnamefont{S.~B.} \bibnamefont{Yuste}},
  \bibinfo{journal}{J. Chem. Phys.} \textbf{\bibinfo{volume}{122}},
  \bibinfo{pages}{024514} (\bibinfo{year}{2005}).

\bibitem[{\citenamefont{Lebowitz}(1964)}]{L64}
\bibinfo{author}{\bibfnamefont{J.~L.} \bibnamefont{Lebowitz}},
  \bibinfo{journal}{Phys. Rev.} \textbf{\bibinfo{volume}{133}},
  \bibinfo{pages}{A895} (\bibinfo{year}{1964}).

\bibitem[{\citenamefont{Yuste et~al.}(1998)\citenamefont{Yuste, Santos, and
  {L\'opez de Haro}}}]{YSH98}
\bibinfo{author}{\bibfnamefont{S.~B.} \bibnamefont{Yuste}},
  \bibinfo{author}{\bibfnamefont{A.}~\bibnamefont{Santos}}, \bibnamefont{and}
  \bibinfo{author}{\bibfnamefont{M.}~\bibnamefont{{L\'opez de Haro}}},
  \bibinfo{journal}{J. Chem. Phys.} \textbf{\bibinfo{volume}{108}},
  \bibinfo{pages}{3683} (\bibinfo{year}{1998}).

\bibitem[{\citenamefont{Rohrmann and Santos}(2011)}]{RS11}
\bibinfo{author}{\bibfnamefont{R.~D.} \bibnamefont{Rohrmann}} \bibnamefont{and}
  \bibinfo{author}{\bibfnamefont{A.}~\bibnamefont{Santos}},
  \bibinfo{journal}{Phys. Rev. E} \textbf{\bibinfo{volume}{83}},
  \bibinfo{pages}{011201} (\bibinfo{year}{2011}).

\bibitem[{\citenamefont{Salsburg et~al.}(1953)\citenamefont{Salsburg, Zwanzig,
  and Kirkwood}}]{SZK53}
\bibinfo{author}{\bibfnamefont{Z.~W.} \bibnamefont{Salsburg}},
  \bibinfo{author}{\bibfnamefont{R.~W.} \bibnamefont{Zwanzig}},
  \bibnamefont{and} \bibinfo{author}{\bibfnamefont{J.~G.}
  \bibnamefont{Kirkwood}}, \bibinfo{journal}{J. Chem. Phys.}
  \textbf{\bibinfo{volume}{21}}, \bibinfo{pages}{1098} (\bibinfo{year}{1953}).

\bibitem[{\citenamefont{Lebowitz and Zomick}(1971)}]{LZ71}
\bibinfo{author}{\bibfnamefont{J.~L.} \bibnamefont{Lebowitz}} \bibnamefont{and}
  \bibinfo{author}{\bibfnamefont{D.}~\bibnamefont{Zomick}},
  \bibinfo{journal}{J. Chem. Phys.} \textbf{\bibinfo{volume}{54}},
  \bibinfo{pages}{3335} (\bibinfo{year}{1971}).

\bibitem[{\citenamefont{Heying and Corti}(2004)}]{HC04}
\bibinfo{author}{\bibfnamefont{M.}~\bibnamefont{Heying}} \bibnamefont{and}
  \bibinfo{author}{\bibfnamefont{D.~S.} \bibnamefont{Corti}},
  \bibinfo{journal}{Fluid Phase Equil.} \textbf{\bibinfo{volume}{220}},
  \bibinfo{pages}{85} (\bibinfo{year}{2004}).

\bibitem[{\citenamefont{Santos}(2007)}]{S07}
\bibinfo{author}{\bibfnamefont{A.}~\bibnamefont{Santos}},
  \bibinfo{journal}{Phys. Rev. E} \textbf{\bibinfo{volume}{76}},
  \bibinfo{pages}{062201} (\bibinfo{year}{2007}).

\bibitem[{\citenamefont{Widom and Rowlinson}(1970)}]{WR70}
\bibinfo{author}{\bibfnamefont{B.}~\bibnamefont{Widom}} \bibnamefont{and}
  \bibinfo{author}{\bibfnamefont{J.}~\bibnamefont{Rowlinson}},
  \bibinfo{journal}{J. Chem. Phys.} \textbf{\bibinfo{volume}{15}},
  \bibinfo{pages}{1670} (\bibinfo{year}{1970}).

\bibitem[{\citenamefont{Ruelle}(1971)}]{R71}
\bibinfo{author}{\bibfnamefont{D.}~\bibnamefont{Ruelle}},
  \bibinfo{journal}{Phys. Rev. Lett.} \textbf{\bibinfo{volume}{16}},
  \bibinfo{pages}{1040} (\bibinfo{year}{1971}).

\bibitem[{\citenamefont{Asakura and Oosawa}(1954)}]{AO54}
\bibinfo{author}{\bibfnamefont{S.}~\bibnamefont{Asakura}} \bibnamefont{and}
  \bibinfo{author}{\bibfnamefont{F.}~\bibnamefont{Oosawa}},
  \bibinfo{journal}{J. Chem. Phys.} \textbf{\bibinfo{volume}{22}},
  \bibinfo{pages}{1255} (\bibinfo{year}{1954}).

\bibitem[{\citenamefont{Asakura and Oosawa}(1958)}]{AO58}
\bibinfo{author}{\bibfnamefont{S.}~\bibnamefont{Asakura}} \bibnamefont{and}
  \bibinfo{author}{\bibfnamefont{F.}~\bibnamefont{Oosawa}},
  \bibinfo{journal}{J. Polym. Sci.} \textbf{\bibinfo{volume}{33}},
  \bibinfo{pages}{183} (\bibinfo{year}{1958}).

\bibitem[{\citenamefont{Rovere and Pastore}(1994)}]{RP94}
\bibinfo{author}{\bibfnamefont{M.}~\bibnamefont{Rovere}} \bibnamefont{and}
  \bibinfo{author}{\bibfnamefont{G.}~\bibnamefont{Pastore}},
  \bibinfo{journal}{J. Phys.: Condens. Matter} \textbf{\bibinfo{volume}{6}},
  \bibinfo{pages}{A163} (\bibinfo{year}{1994}).

\bibitem[{\citenamefont{Jagannathan and Yethiraj}(2003)}]{JY03}
\bibinfo{author}{\bibfnamefont{K.}~\bibnamefont{Jagannathan}} \bibnamefont{and}
  \bibinfo{author}{\bibfnamefont{A.}~\bibnamefont{Yethiraj}},
  \bibinfo{journal}{J. Chem. Phys.} \textbf{\bibinfo{volume}{118}},
  \bibinfo{pages}{7907} (\bibinfo{year}{2003}).

\bibitem[{\citenamefont{G\'o\'zd\'z}(2003)}]{G03}
\bibinfo{author}{\bibfnamefont{W.~T.} \bibnamefont{G\'o\'zd\'z}},
  \bibinfo{journal}{J. Chem. Phys.} \textbf{\bibinfo{volume}{119}},
  \bibinfo{pages}{3309} (\bibinfo{year}{2003}).

\bibitem[{\citenamefont{Buhot}(2005)}]{B05}
\bibinfo{author}{\bibfnamefont{A.}~\bibnamefont{Buhot}}, \bibinfo{journal}{J.
  Chem. Phys.} \textbf{\bibinfo{volume}{122}}, \bibinfo{pages}{024105}
  (\bibinfo{year}{2005}).

\bibitem[{\citenamefont{Lomba et~al.}(1996)\citenamefont{Lomba, Alvarez, Lee,
  and Almarza}}]{LALA96}
\bibinfo{author}{\bibfnamefont{E.}~\bibnamefont{Lomba}},
  \bibinfo{author}{\bibfnamefont{M.}~\bibnamefont{Alvarez}},
  \bibinfo{author}{\bibfnamefont{L.~L.} \bibnamefont{Lee}}, \bibnamefont{and}
  \bibinfo{author}{\bibfnamefont{N.~G.} \bibnamefont{Almarza}},
  \bibinfo{journal}{J. Chem. Phys.} \textbf{\bibinfo{volume}{104}},
  \bibinfo{pages}{4180} (\bibinfo{year}{1996}).

\bibitem[{\citenamefont{Santos and {L\'opez de Haro}}(2005)}]{SH05}
\bibinfo{author}{\bibfnamefont{A.}~\bibnamefont{Santos}} \bibnamefont{and}
  \bibinfo{author}{\bibfnamefont{M.}~\bibnamefont{{L\'opez de Haro}}},
  \bibinfo{journal}{Phy. Rev. E} \textbf{\bibinfo{volume}{72}},
  \bibinfo{pages}{010501(R)} (\bibinfo{year}{2005}).

\bibitem[{\citenamefont{Sillr\'en and Hansen}(2010)}]{SH10}
\bibinfo{author}{\bibfnamefont{P.}~\bibnamefont{Sillr\'en}} \bibnamefont{and}
  \bibinfo{author}{\bibfnamefont{J.-P.} \bibnamefont{Hansen}},
  \bibinfo{journal}{Mol. Phys.} \textbf{\bibinfo{volume}{105}},
  \bibinfo{pages}{1803} (\bibinfo{year}{2010}).

\bibitem[{\citenamefont{Yuste et~al.}(2008)\citenamefont{Yuste, Santos, and
  {L\'opez de Haro}}}]{YSH08}
\bibinfo{author}{\bibfnamefont{S.~B.} \bibnamefont{Yuste}},
  \bibinfo{author}{\bibfnamefont{A.}~\bibnamefont{Santos}}, \bibnamefont{and}
  \bibinfo{author}{\bibfnamefont{M.}~\bibnamefont{{L\'opez de Haro}}},
  \bibinfo{journal}{J. Chem. Phys.} \textbf{\bibinfo{volume}{128}},
  \bibinfo{pages}{134507} (\bibinfo{year}{2008}).

\bibitem[{\citenamefont{Lebowitz et~al.}(1962)\citenamefont{Lebowitz, Percus,
  and Zucker}}]{LPZ62}
\bibinfo{author}{\bibfnamefont{J.~L.} \bibnamefont{Lebowitz}},
  \bibinfo{author}{\bibfnamefont{J.~K.} \bibnamefont{Percus}},
  \bibnamefont{and} \bibinfo{author}{\bibfnamefont{I.~J.}
  \bibnamefont{Zucker}}, \bibinfo{journal}{Bull. Am. Phys. Soc.}
  \textbf{\bibinfo{volume}{7}}, \bibinfo{pages}{415} (\bibinfo{year}{1962}).

\bibitem[{\citenamefont{Ben-Naim and Santos}(2009)}]{BNS09}
\bibinfo{author}{\bibfnamefont{A.}~\bibnamefont{Ben-Naim}} \bibnamefont{and}
  \bibinfo{author}{\bibfnamefont{A.}~\bibnamefont{Santos}},
  \bibinfo{journal}{J. Chem. Phys.} \textbf{\bibinfo{volume}{131}},
  \bibinfo{pages}{164512} (\bibinfo{year}{2009}).

\bibitem[{\citenamefont{{L\'opez de Haro} et~al.}(2008)\citenamefont{{L\'opez
  de Haro}, Yuste, and Santos}}]{HYS08}
\bibinfo{author}{\bibfnamefont{M.}~\bibnamefont{{L\'opez de Haro}}},
  \bibinfo{author}{\bibfnamefont{S.~B.} \bibnamefont{Yuste}}, \bibnamefont{and}
  \bibinfo{author}{\bibfnamefont{A.}~\bibnamefont{Santos}}, in
  \emph{\bibinfo{booktitle}{{Theory and Simulation of Hard-Sphere Fluids and
  Related Systems}}}, edited by
  \bibinfo{editor}{\bibfnamefont{A.}~\bibnamefont{Mulero}}
  (\bibinfo{publisher}{Springer}, \bibinfo{address}{Berlin},
  \bibinfo{year}{2008}), vol. \bibinfo{volume}{753} of
  \emph{\bibinfo{series}{Lectures Notes in Physics}}, pp.
  \bibinfo{pages}{183--245}.

\bibitem[{\citenamefont{Schmidt}(2007)}]{S07b}
\bibinfo{author}{\bibfnamefont{M.}~\bibnamefont{Schmidt}},
  \bibinfo{journal}{Phys. Rev. E} \textbf{\bibinfo{volume}{76}},
  \bibinfo{pages}{031202} (\bibinfo{year}{2007}).

\bibitem[{\citenamefont{Jung et~al.}(1994{\natexlab{a}})\citenamefont{Jung,
  Jhon, and Ree}}]{JJR94a}
\bibinfo{author}{\bibfnamefont{J.}~\bibnamefont{Jung}},
  \bibinfo{author}{\bibfnamefont{M.~S.} \bibnamefont{Jhon}}, \bibnamefont{and}
  \bibinfo{author}{\bibfnamefont{F.~H.} \bibnamefont{Ree}},
  \bibinfo{journal}{J. Chem. Phys.} \textbf{\bibinfo{volume}{100}},
  \bibinfo{pages}{528} (\bibinfo{year}{1994}{\natexlab{a}}).

\bibitem[{\citenamefont{Jung et~al.}(1994{\natexlab{b}})\citenamefont{Jung,
  Jhon, and Ree}}]{JJR94b}
\bibinfo{author}{\bibfnamefont{J.}~\bibnamefont{Jung}},
  \bibinfo{author}{\bibfnamefont{M.~S.} \bibnamefont{Jhon}}, \bibnamefont{and}
  \bibinfo{author}{\bibfnamefont{F.~H.} \bibnamefont{Ree}},
  \bibinfo{journal}{J. Chem. Phys.} \textbf{\bibinfo{volume}{100}},
  \bibinfo{pages}{9064} (\bibinfo{year}{1994}{\natexlab{b}}).

\bibitem[{\citenamefont{Hamad}(1997)}]{H97}
\bibinfo{author}{\bibfnamefont{E.~Z.} \bibnamefont{Hamad}},
  \bibinfo{journal}{Mol. Phys.} \textbf{\bibinfo{volume}{91}},
  \bibinfo{pages}{371} (\bibinfo{year}{1997}).

\bibitem[{\citenamefont{Boubl\'{\i}k}(1970)}]{B70}
\bibinfo{author}{\bibfnamefont{T.}~\bibnamefont{Boubl\'{\i}k}},
  \bibinfo{journal}{J. Chem. Phys.} \textbf{\bibinfo{volume}{53}},
  \bibinfo{pages}{471} (\bibinfo{year}{1970}).

\bibitem[{\citenamefont{Mansoori et~al.}(1971)\citenamefont{Mansoori, Carnahan,
  and K.~E. Starlingand T. W.~Leland}}]{MCSL71}
\bibinfo{author}{\bibfnamefont{G.~A.} \bibnamefont{Mansoori}},
  \bibinfo{author}{\bibfnamefont{N.~F.} \bibnamefont{Carnahan}},
  \bibnamefont{and} \bibinfo{author}{\bibfnamefont{J.}~\bibnamefont{K.~E.
  Starlingand T. W.~Leland}}, \bibinfo{journal}{J. Chem. Phys.}
  \textbf{\bibinfo{volume}{54}}, \bibinfo{pages}{1523} (\bibinfo{year}{1971}).

\bibitem[{\citenamefont{Grundke and Henderson}(1972)}]{GH72}
\bibinfo{author}{\bibfnamefont{E.~W.} \bibnamefont{Grundke}} \bibnamefont{and}
  \bibinfo{author}{\bibfnamefont{D.}~\bibnamefont{Henderson}},
  \bibinfo{journal}{Mol. Phys.} \textbf{\bibinfo{volume}{24}},
  \bibinfo{pages}{269} (\bibinfo{year}{1972}).

\bibitem[{\citenamefont{Lee and Levesque}(1973)}]{LL73}
\bibinfo{author}{\bibfnamefont{L.~L.} \bibnamefont{Lee}} \bibnamefont{and}
  \bibinfo{author}{\bibfnamefont{D.}~\bibnamefont{Levesque}},
  \bibinfo{journal}{Mol. Phys.} \textbf{\bibinfo{volume}{26}},
  \bibinfo{pages}{1351} (\bibinfo{year}{1973}).

\bibitem[{\citenamefont{Fantoni and Pastore}(2004)}]{FP04}
\bibinfo{author}{\bibfnamefont{R.}~\bibnamefont{Fantoni}} \bibnamefont{and}
  \bibinfo{author}{\bibfnamefont{G.}~\bibnamefont{Pastore}},
  \bibinfo{journal}{Physics A} \textbf{\bibinfo{volume}{332}},
  \bibinfo{pages}{349} (\bibinfo{year}{2004}), \bibinfo{note}{{Note that there
  is a misprint in Eq.\ (13), which should read
  $\bar{h}_{12}(k)=\bar{c}_{12}(k)[1-\rho_1\rho_2\bar{c}_{12}^2(k)]^{-1}$}}.

\bibitem[{\citenamefont{Abate and Whitt}(1992)}]{AW92}
\bibinfo{author}{\bibfnamefont{J.}~\bibnamefont{Abate}} \bibnamefont{and}
  \bibinfo{author}{\bibfnamefont{W.}~\bibnamefont{Whitt}},
  \bibinfo{journal}{Queueing Systems} \textbf{\bibinfo{volume}{10}},
  \bibinfo{pages}{5} (\bibinfo{year}{1992}).

\bibitem[{not()}]{note_11_09}
\bibinfo{note}{See Supplemental Material at
  \url{http://link.aps.org/supplemental/10.1103/PhysRevE.84.041201} for a
  Mathematica notebook with a code to evaluate $g_{ij}(r)$ from approximation
  $\text{RFA}_+^{(1)}$. The notebook can also be downloaded from
  [\url{http://www.unex.es/eweb/fisteor/andres/NAHS/gij_NAHS.nb}].}

\bibitem[{\citenamefont{Allen and Tildesley}(1987)}]{AT87}
\bibinfo{author}{\bibfnamefont{M.~P.} \bibnamefont{Allen}} \bibnamefont{and}
  \bibinfo{author}{\bibfnamefont{D.~J.} \bibnamefont{Tildesley}},
  \emph{\bibinfo{title}{{Computer Simulation of Liquids}}}
  (\bibinfo{publisher}{Clarendon Press}, \bibinfo{address}{Oxford},
  \bibinfo{year}{1987}).

\bibitem[{\citenamefont{Shew and Yethiraj}(1996)}]{SY96}
\bibinfo{author}{\bibfnamefont{C.-Y.} \bibnamefont{Shew}} \bibnamefont{and}
  \bibinfo{author}{\bibfnamefont{A.}~\bibnamefont{Yethiraj}},
  \bibinfo{journal}{J. Chem. Phys.} \textbf{\bibinfo{volume}{104}},
  \bibinfo{pages}{7665} (\bibinfo{year}{1996}).

\bibitem[{\citenamefont{Johnson et~al.}(1997)\citenamefont{Johnson, Gould,
  Machta, and Chayes}}]{JGMC97}
\bibinfo{author}{\bibfnamefont{G.}~\bibnamefont{Johnson}},
  \bibinfo{author}{\bibfnamefont{H.}~\bibnamefont{Gould}},
  \bibinfo{author}{\bibfnamefont{J.}~\bibnamefont{Machta}}, \bibnamefont{and}
  \bibinfo{author}{\bibfnamefont{L.~K.} \bibnamefont{Chayes}},
  \bibinfo{journal}{Phys. Rev. Lett.} \textbf{\bibinfo{volume}{79}},
  \bibinfo{pages}{2612} (\bibinfo{year}{1997}).

\end{thebibliography}
\end{document}